\documentclass[A4paper,authoryear,11pt]{elsarticle}
\usepackage{amssymb}

\usepackage{amssymb}
\usepackage{amsmath}
\usepackage{lipsum}
\usepackage{gensymb}

\usepackage{algorithm}
\usepackage{algpseudocodex}
\usepackage{hyperref}
\usepackage{booktabs}
\usepackage{multirow}
\usepackage{multicol}
\usepackage{graphicx}
\usepackage{adjustbox}
\usepackage{makecell}
\usepackage{capt-of}
\usepackage{threeparttable}

\hypersetup{
    colorlinks=true,
    linkcolor=blue,
    citecolor=blue,
    urlcolor=blue,
}

\begin{document}

\begin{frontmatter}



\title{MergeDJD: A Fast Constructive Algorithm with Piece Merging for the Two-Dimensional Irregular Bin Packing Problem}
 \author[uestc,cuhk]{Yiping Liu}\ead{liuyiping@cuhk.edu.cn}
 \affiliation[uestc]{organization={School of Computer Science and Engineering, University of Electronic Science and Technology of China},
             city={Chengdu},
             postcode={611731}, 
             country={China}}            
 \author[uestc]{Haocheng Fu}\ead{hfu622@std.uestc.edu.cn}

 \author[uestc]{Yi Zhou\corref{cor1}}\ead{zhou.yi@uestc.edu.cn}
 \author[uestc]{Jian Mao}\ead{maojian@std.uestc.edu.cn}

 \author[cuhk]{Zhang-Hua Fu}\ead{fuzhanghua@cuhk.edu.cn}
\affiliation[cuhk]{organization={School of Data Science {\protect\upshape\&} School of Science and Engineering, The Chinese University of Hong Kong},
    city={Shenzhen},
    postcode={518129},
    country={China}}
  
 \cortext[cor1]{Corresponding author}
             
\begin{abstract}
The two-dimensional irregular bin packing problem (2DIBPP) aims to pack a given set of irregular polygons, referred to as pieces, into fixed-size rectangular bins without overlap, while maximizing bin utilization.
Although numerous metaheuristic algorithms have been proposed for the 2DIBPP, many industrial applications favor simpler constructive heuristics due to their deterministic behavior and low computational overhead.
Among such methods, the DJD algorithm proposed by López-Camacho et al. is one of the most competitive constructive heuristics for the 2DIBPP. 
However, DJD is less effective for cutting instances, in which many pieces can be seamlessly combined into larger polygons.
To address the issue, we propose MergeDJD, a novel constructive algorithm that integrates and extends the DJD framework. 
MergeDJD first preprocesses the instance by iteratively identifying groups of pieces that can be combined into larger and more regular pieces. 
It then employs an improved version of DJD, in which the placement strategy is enhanced to better handle non-convex and combined shapes, to pack all resulting pieces into bins.
Computational experiments on 1,089 well-known benchmark instances show that MergeDJD consistently outperforms DJD on 1,083 instances while maintaining short runtimes. Notably, MergeDJD attains new best known values on 515 instances. 
Ablation studies further confirm the effectiveness of the proposed components. To facilitate reproducibility and future research, we have open-sourced the complete implementation and provided interfaces for visualizing packing results.
\end{abstract}



\end{frontmatter}



\section{Introduction}
\label{Intro}


The two-dimensional irregular bin packing problem (2DIBPP) is a well-known and challenging industrial combinatorial optimization problem. 
In general, the problem asks to place a given set of irregular polygonal pieces into multiple rectangular bins without overlap, while maximizing the bin utilization. 
The 2DIBPP is of significant importance in industry, particularly in nesting processes where two-dimensional parts are cut from rectangular sheets \citep{francescatto2025two}. Typical applications can be found in the manufacturing of cars and ships \citep{zhang2022iteratively,sari2025systematic}, the garment industry \citep{Okano}, and the glass production \citep{Martinez-Sykora2017}. 
In these settings, high-quality packing solutions directly reduce material waste and provide substantial economic benefits.


However, solving the 2DIBPP is very challenging both theoretically and practically. From a theoretical perspective, the problem remains NP-hard even when all pieces are restricted to rectangles (i.e., 2D bin packing) \citep{lodi2002two} or in a one-dimensional setting (i.e, traditional bin packing) \citep{korf2002new}. 
In practice, exact solution approaches are only applicable to very small instances, which limits their usefulness in real-world applications. 
As a result, a wide range of heuristic approaches have been developed to obtain high-quality packing within a reasonable computational time.

\subsection{Literature Review}
\label{review}
In this section, we summarize the existing heuristic algorithms. 
In general, these approaches can be broadly classified into two categories: \emph{constructive algorithms} and \emph{improvement algorithms}.

\paragraph{Constructive algorithms} 
A constructive algorithm typically starts from an empty solution and incrementally extends a partial solution until a complete feasible solution is obtained.
By making irrevocable decisions, constructive heuristics can generate feasible solutions with low computational overhead.
For the 2DIBPP, there are two heuristic decisions need to be made in the constructive algorithm, the \emph{piece-to-bin assignment} and \emph{in-bin positioning}. 
The former determines to which bin a piece is assigned, while the latter decides the exact placement position within the selected bin. 

For the piece-to-bin assignment stage, \emph{First Fit Decreasing} (FFD) sorts pieces in decreasing order of area and assigns each piece to the first bin in which it can be feasibly placed; if no such bin exists, a new bin is opened. 
\emph{Best Fit Decreasing} (BFD) instead selects the bin that results in the minimum residual free space after placement. Within each bin, pieces are commonly processed in decreasing order of area. Both FFD and BFD are originally introduced for packing rectangle pieces \citep{terashima2006ga}. 
They are adapted for irregular pieces by \citet{López-Camacho2013}. Based on these principles, the DJD heuristic proposed in \citet{López-Camacho2013} further refines the placement order and bin assignment to improve bin utilization.



For the in-bin positioning stage, the \emph{bottom-left} heuristic places a piece at the lowest feasible position and, ties breaking with the leftmost one \citep{dowsland2002algorithm}.
The \emph{maximum utilization} heuristic selects positions that maximize area usage in the earliest bin, whereas the \emph{minimum length} heuristic aims to minimize the length of the enclosing rectangle of the partial solution \citep{burke2006new}. 
The CAD heuristic further improves placement quality by maximizing adjacency between the placed piece and existing pieces or bin boundaries \citep{uday2001nesting}.


Due to their efficiency and simplicity, constructive heuristics are widely used either as standalone methods or as initialization components for more sophisticated improvement algorithms.


\paragraph{Improvement algorithms}

Unlike constructive heuristics that generate a solution by making a sequence of irrevocable  decisions, \emph{improvement algorithms} employ an iterative search process that repeatedly applies heuristic rules with a richer computational budget. 
In this perspective, two representative classes of improvement algorithms for the 2DIBPP are \emph{local search} and \emph{hyper-heuristic}.


Local search starts from a feasible solution (which is sometimes generated by a constructive heuristic), repeatedly performs local modifications, like perturbing placements within a bin or exchanging pieces across bins, that lead to better solutions. 
To be specific, \citet{Martinez-Sykora2017} formulated bin assignment and packing using integer and mixed-integer programming models, combined with local improvement mechanisms. 
\citet{abeysooriya2018jostle} proposed the \emph{Jostle} heuristic, which improves an initial placement by locally perturbing piece positions and orientations. 
\citet{Liu2020,zhang2022iteratively} generated initial solutions via FFD and bottom-left heuristics and then applied relocation and exchange operations across bins. 
More recent studies explored alternative representations and neighborhood designs, including raster-based modeling, overlap minimization, and iterative space contraction \citep{luo2025decimal,yao2024iteratively,tang2024iterative}. 
In addition, \citet{wang2022optimization,cai2023heuristics} improved bin utilization by decomposing the problem and applying local improvement and repacking strategies.

Hyper-heuristic methods, on the other hand, conduct search at the level of heuristics rather than directly manipulating a single packing. 
They iteratively select, adapt, or combine low-level heuristics (often constructive rules and local improvement moves) to generate a sequence of candidate solutions, with the selection guided by feedback from solution quality \citep{Terashima-Marín2010,lopez2014unified,Guerriero}. 
Genetic algorithm–based hyper-heuristics were proposed in \citet{Terashima-Marín2010,lopez2014unified} to evolve effective heuristic combinations for the 2DIBPP. More recently, \citet{Guerriero} introduced a dynamic and hierarchical hyper-heuristic framework, where heuristic selection is guided by instance characteristics and may recursively invoke either simple heuristics or other hyper-heuristics.

\subsection{Contribution and Organization}
\label{subsec contributions organization}

As already pointed out by \citet{López-Camacho2013}, practitioners often favor simple, deterministic, and transparent algorithms over black-box methods with long runtimes, although the latter may produce solutions of higher quality.
As a result, simple constructive heuristics are attractive for time- and cost-sensitive industrial applications due to their predictable behavior and low computational overhead.
Motivated by this practical preference, in this paper, we continue to work on constructive algorithms for the 2DBIPP by making the following contributions.

\begin{itemize}
    \item 
    We identify a key limitation of existing constructive heuristics on cutting instances in which pieces exhibit strong geometric compatibility. 
    Based on this, we propose a merge strategy that identifies compatible pieces and combines them into larger and more regular super-shapes.
    \item   
    We develop a new constructive heuristic algorithm for the 2DIBPP, namely \emph{MergeDJD}, which substantially extends the DJD heuristic \citep{López-Camacho2013}.
    Firstly, MergeDJD integrates the proposed merge strategy as a preprocessing stage.
    Moreover, MergeDJD incorporates an improved placement heuristic (ICAD) to better handle non-convex and combined shapes, while it still preserves  the deterministic and low-overhead of constructive algorithms.
    \item 
    We conduct extensive computational experiments on 1,089 benchmark instances. 
    MergeDJD consistently outperforms DJD and attains new best known values on 515 instances. 
    Ablation studies further validate the effectiveness of the proposed components. 
    The complete implementation with visualization support is publicly available at \url{https://github.com/hfu622/2DIBPP-MergeDJD}.
\end{itemize}


This paper is organized as follows. 
Section~\ref{description} provides a detailed description of the 2DIBPP. 
In Section~\ref{sectDJD++}, we revisit the DJD and point out its limitation, then we introduce the overall framework of the proposed MergeDJD. 
Then, in Sections~\ref{sec:merge} and~\ref{CAD}, we describe in detail the two main components of MergeDJD, a piece-merge algorithm and an ICAD placement heuristic (within the DJD), respectively. 
Section~\ref{result} reports the experimental results along with comprehensive comparisons and analyses. 
Finally, Section~\ref{sec:conclusion} concludes the paper.

\section{Problem Description}
\label{description}

We formally define the two-dimensional irregular bin packing problem (2DIBPP) and show how the geometric data are represented in this problem.

The input of the 2DIBPP is a tuple $(\mathcal{P}, W, L, \mathcal{A})$, where
$\mathcal{P}=\{P_1,P_2,\ldots,P_n\}$ is a set of $n$ pieces,
$W,L\in\mathbb{R}^+$ denote the width and length of each bin, and
$\mathcal{A}\subseteq[0,2\pi)$ is a finite set of allowed rotation angles.

\begin{figure}[h!]
    \centering
    \includegraphics[width=0.4\linewidth]{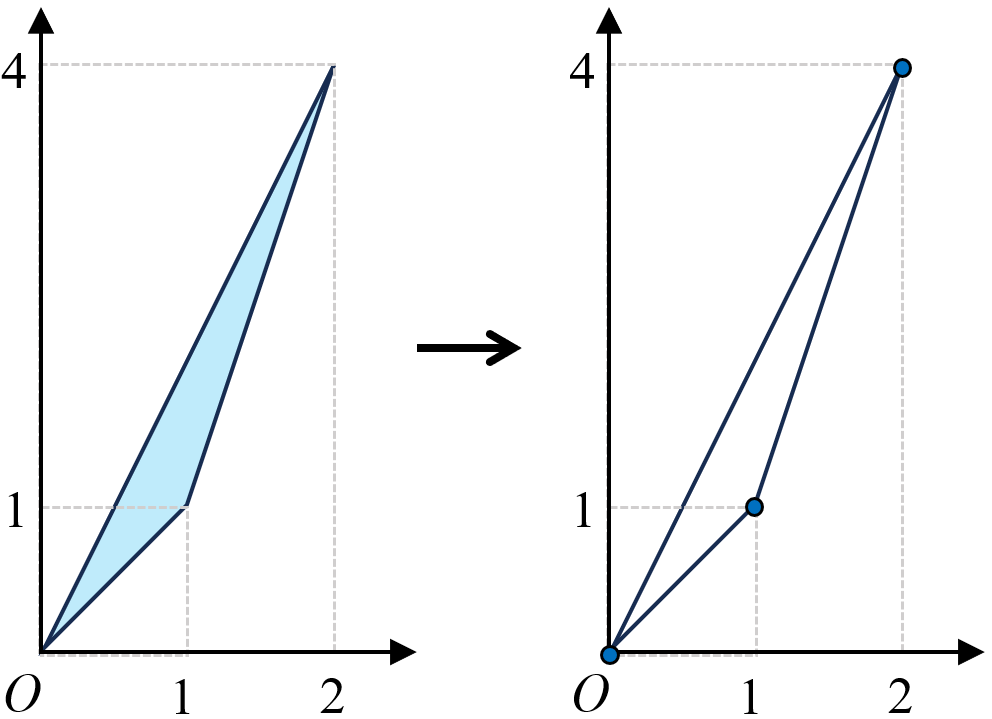}
    \caption{The triangle piece can be represented by $\langle (0,0),(1,1),(2,4)\rangle$.}
    \label{fig:triangle}
\end{figure}

We assume that each input piece $P\in\mathcal{P}$ is a simple polygon without holes, represented by an ordered sequence of two-dimensional vertices.
Without loss of generality, the first vertex of each piece is fixed at $(0,0)$ and is referred to as the \emph{reference point}.
Vertices are given in either clockwise or counterclockwise order.
For example, if a piece is a triangle as shown in Figure~\ref{fig:triangle}, it can be represented as $\langle(0,0),(1,1),(2,4)\rangle$.
Each bin is represented by the sequence of corner points $\langle(0,0),(L,0),(L,W),(0,W)\rangle$.
We assume that each piece can be placed into a bin under at least one allowed rotation.
Moreover, we assume an upper bound $b$ on the number of available bins, with $b\ge n$.



The 2DIBPP requires a feasible and cost-effective placement of the pieces in $\mathcal{P}$.
The placement of a subset of pieces $\mathcal{P}'\subseteq\mathcal{P}$ consists of two components: the \textit{assignment} of each piece to a bin, and the \textit{layout} of pieces within each bin.
The layout of a piece specifies both its translation and its rotation angle $\alpha\in\mathcal{A}$.
A placement is feasible if each piece is assigned to exactly one bin, all pieces lie entirely within the boundaries of their assigned bins, and no two pieces assigned to the same bin overlap.

\begin{figure}[h!]
    \centering
    \includegraphics[width=0.6\linewidth]{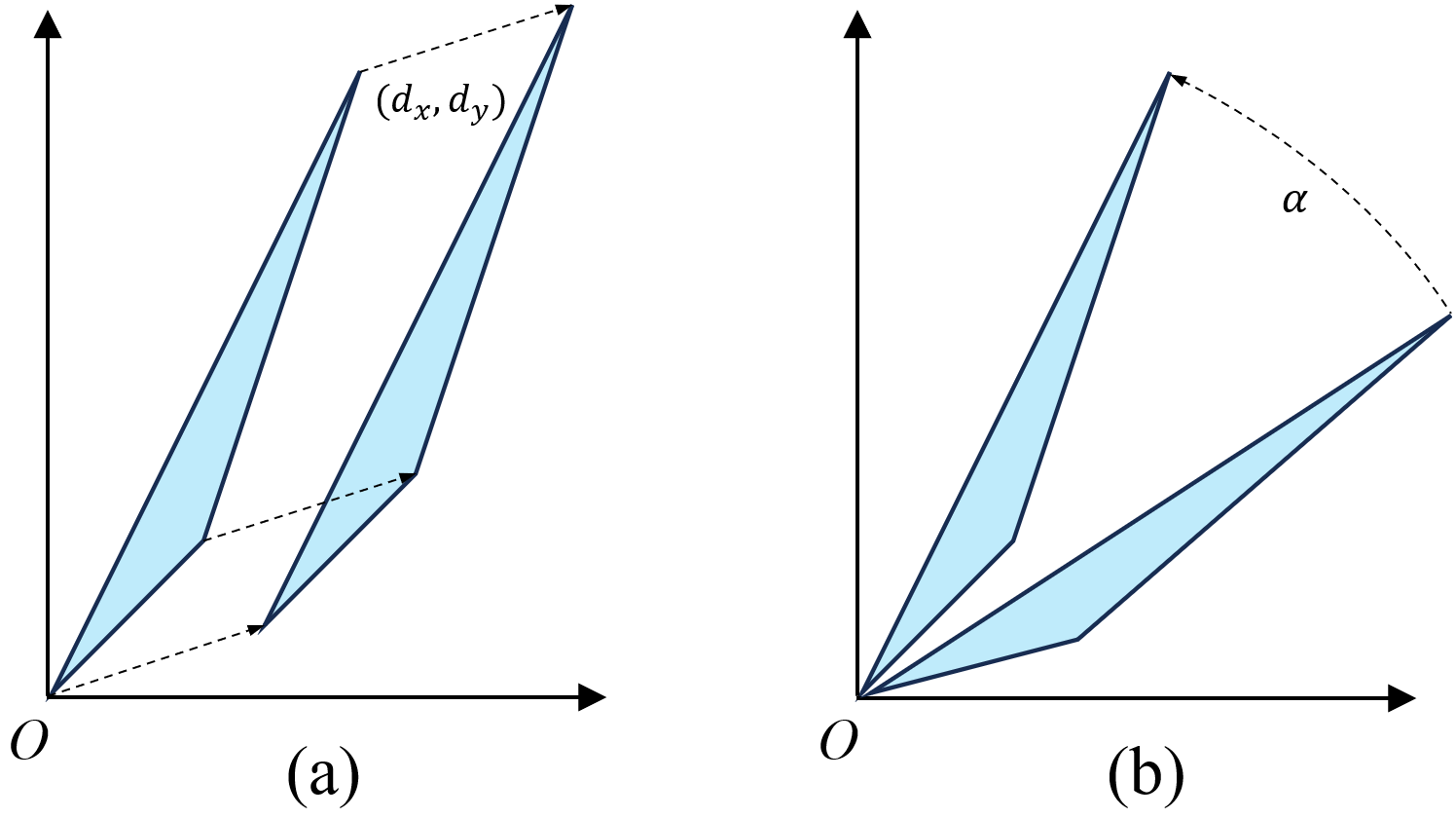}
    \caption{(a) All points of a piece are translated by $(d_x,d_y)$. (b) The piece is rotated counterclockwise by $\alpha$ degrees around its reference point $O$.}
    \label{fig:trans_rot_new}
\end{figure}

To represent a placement, we index the bins as $[b]=\{1,\ldots,b\}$, where $b$ is an upper bound on the number of available bins.
We define a \textit{position triplet} of a piece $P\in\mathcal{P}$ as $v_P=(i,\vec{v},\alpha)$, where $i\in[b]$ indicates that $P$ is placed in the $i$-th bin, $\vec{v}$ is the translation (move) vector, and $\alpha\in\mathcal{A}$ is the rotation angle.
An example illustrating the effect of translation and rotation is shown in Figure~\ref{fig:trans_rot_new}.

A \textit{placement} of a subset $\mathcal{P}'\subseteq\mathcal{P}$ is represented by a set of position triplets $V$, such that for each $P\in\mathcal{P}'$ there exists a unique $v_P=(i,\vec{v},\alpha)\in V$.
If $\mathcal{P}'\subset\mathcal{P}$, the placement is \textit{partial}; if $\mathcal{P}'=\mathcal{P}$, the placement is \textit{complete}.
Since all bins are identical and interchangeable, we assume without loss of generality that bins are used consecutively: bin $i$ $(1<i\le b)$ can appear in a placement only if bin $i-1$ is also used.
We denote by $N$ the number of bins used in a complete placement.


\subsection{The Metric of Placement}
\label{subsec_metric_definition}
Let $Area(P)\in\mathbb{R}^+$ denote the area of a piece $P$, which can be computed using standard polygon area formulas \citep{preparata2012computational}.
Given a feasible placement, let $\mathcal{P}_i$ be the set of pieces assigned to the $i$-th bin.
The \textit{utilization ratio} of bin $i$ is defined as
\[
U_i=\frac{\sum_{P\in\mathcal{P}_i} Area(P)}{W\times L}.
\]

The primary objective of the 2DIBPP is to minimize the number of bins used, i.e., $N$.
In addition, several metrics are commonly used to evaluate the quality of a complete and feasible placement, especially when comparing solutions that use the same number of bins.

\begin{figure}
    \centering
    \includegraphics[width=0.6\linewidth]{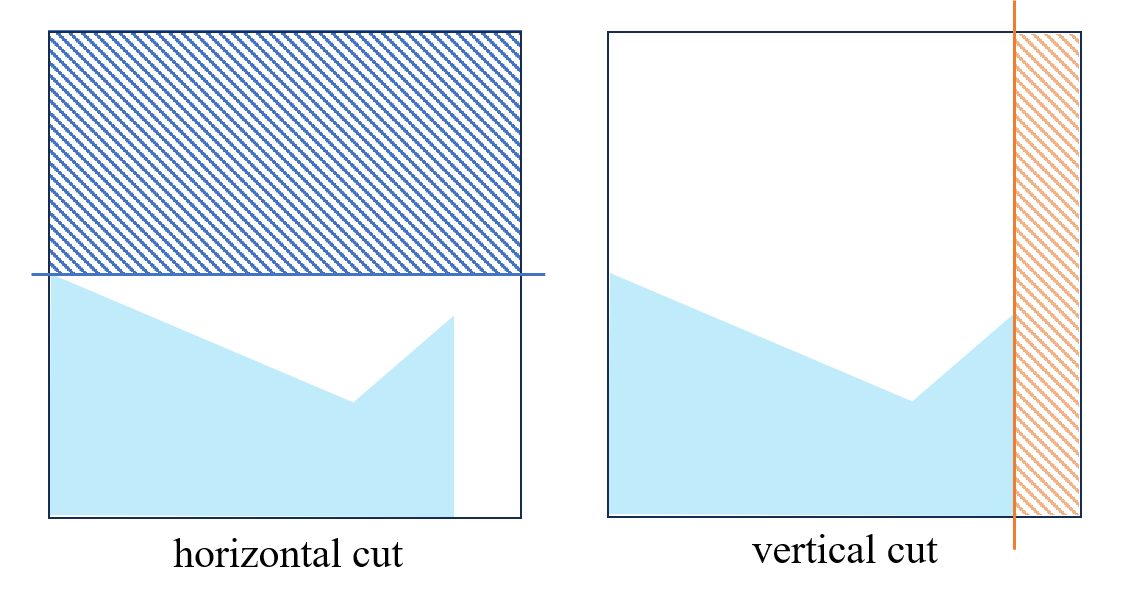}
    \caption{Illustration of horizontal cut and vertical cut. The shadow area represents the reusable residual.}
    \label{fig:K value}
\end{figure}

\begin{itemize}
    \item \textit{The $F$ metric.}  
    This metric is defined as
    \[
    F=\frac{\sum_{i=1}^{N} U_i^2}{N},
    \]
    where $U_i$ is the utilization ratio of bin $i$.
    Given two placements with the same $N$, this metric favors solutions in which bins are either highly utilized or nearly empty.
    The $F$ metric was introduced in \citet{lopez2013understanding} and has been widely used in subsequent studies \citep{zhang2022iteratively}.

    \item \textit{The $K$ metric.}  
    The $K$ metric is defined as
    \[
    K = N - 1 + R^*,
    \]
    where $R^*$ denotes the effective utilization of the least utilized bin.
    Specifically, for the least utilized bin, any completely unoccupied reusable residual area is first removed by applying either a horizontal or vertical guillotine cut, with the cut orientation chosen to maximize the remaining rectangular area.
    The value $R^*$ is then computed as the ratio between the area occupied by pieces in the remaining rectangle and the bin area.
    Thus, $K$ can be interpreted as the number of fully utilized bins plus the fractional utilization of the least utilized bin.
    An example is shown in Figure~\ref{fig:K value}. 
    This metric is particularly useful for distinguishing solutions that use the same number of bins but differ in how efficiently the last bin is packed \citep{han2013construction,zhang2022iteratively}.
\end{itemize}

\section{A MergeDJD Framework for 2DIBPP}
\label{sectDJD++}

\subsection{A Review of the DJD heuristic}
\label{reviewDJD}


In general, the DJD algorithm packs pieces into bins in a greedy, bin-by-bin manner.
All pieces are first sorted in descending order of area.
The algorithm then opens bins sequentially and fills each bin using a two-phase constructive procedure.

For a given bin, DJD first performs a greedy filling phase based on individual pieces.
Following the sorted order, the algorithm repeatedly selects the largest unplaced piece and attempts to place it into the current bin.
Feasibility and exact placement positions are determined by a placement subroutine called the \emph{Constructive Approach–Maximum Adjacency} (CAD) \citep{López-Camacho2013}.
CAD favors placements that maximize adjacency with already placed pieces and bin boundaries.
If a piece can be successfully placed, it is permanently assigned to the current bin.
This process continues until no further piece fits or the occupied area of the bin reaches one third of the bin area, as shown in lines~4–8 of Algorithm~\ref{alg:DJD}.

\begin{algorithm}[ht]
\caption{The DJD algorithm}
\label{alg:DJD}
\textbf{Input:} a set of pieces $\mathcal{P}$, the width $W$ and the length $L$ of each bin. \\
\textbf{Output:} a complete placement $V$.
\begin{algorithmic}[1]
\State  $waste \gets 0$, placed area $\Phi_i\leftarrow 0$ and placement $V_i\leftarrow \emptyset,\forall i\in [1,b]$
\State $i\leftarrow 1$, sort pieces in $\mathcal{P}$ by the descending order of area
\While {$\mathcal{P}\neq \emptyset$}
    \For{$j=1,\ldots,|\mathcal{P}|$}   \Comment{fill the $i$-th bin until one-third}
        \State $flag,v_{P_j}\leftarrow \text{CAD}(i,\langle P_j\rangle)$     
        \If{$flag=$ success}
            \State $\mathcal{P}\leftarrow \mathcal{P}\setminus \{P_j\}$, $\Phi_i\leftarrow \Phi_i+Area(P_j)$, $V_i\leftarrow V_i\cup \{v_{P_j}\}$
        \EndIf
        \If{$\Phi_i\geq W\times L/3$}
            \textbf{break}
        \EndIf
    \EndFor
    
    \While{$waste<W\times L-\Phi_i$}   \Comment{fill the $i$-th bin using groups of pieces}
        \State $\mathcal{G}_1\leftarrow \mathcal{P}$, $\mathcal{G}_2\leftarrow \mathcal{P}\times \mathcal{P}$, $\mathcal{G}_3\leftarrow \mathcal{P}\times \mathcal{P}\times \mathcal{P}$
        \State $flag\leftarrow failure$
        \For{each group of ordered pieces $\vec{O}\in \mathcal{G}_1 \cup \mathcal{G}_2 \cup \mathcal{G}_3$}
            \State $flag,\{v_{P}\mid P\in \vec{O}\}\leftarrow \text{CAD}(i,\vec{O})$
            \If{$flag=$ success and $\Phi_i+\sum_{P\in G}Area(P)\geq W\times L-waste$}
                \State $\mathcal{P}\leftarrow \mathcal{P}\setminus \{P\mid P \in \vec{O}\}$, $\Phi_i\leftarrow \Phi_i+\sum_{P\in \vec{O}}Area(P)$
                \State $V_i\leftarrow V_i\cup \{v_{P}\mid P\in \vec{O}\}$, $waste\leftarrow 0$, \textbf{break}
            \EndIf
        \EndFor
        \If {$flag=$ failure} $waste \gets waste + W\times L/20$
        \EndIf
    \EndWhile
    \State $i\leftarrow i+1$
\EndWhile
\State \Return $V=\{V_1,V_2,\ldots,V_b\}$
\end{algorithmic}
\end{algorithm}

After the initial greedy phase, DJD attempts to further fill the current bin using small groups of remaining pieces.
Specifically, the algorithm considers ordered groups consisting of one, two, or three pieces.
For each group, CAD is invoked to test whether the group can be jointly placed into the bin.
A threshold parameter, denoted by \textit{waste}, controls the acceptable residual empty area.
If a group leads to a sufficiently high utilization according to the current waste threshold, the group is accepted and permanently placed.
If no feasible group is found, the waste threshold is gradually increased, allowing solutions with larger residual space.
This group-based filling process corresponds to lines~9–17 of Algorithm~\ref{alg:DJD}.

Once no additional pieces or groups can be placed, the current bin is closed and a new bin is opened.
The algorithm repeats the same procedure until all pieces have been assigned to bins.
All placement and assignment decisions are made irrevocably.
Therefore, DJD is a constructive heuristic that determines the bin assignment and the layout of pieces within each bin irrevocably.

\subsection{A New Algorithm for the 2DIBPP} 
\label{sec: main framework}

The DJD algorithm is efficient and near-real-time in experiments \citep{López-Camacho2013}.
However, in industrial  applications, some of the input pieces are intermediate components that will later be reassembled by gluing, stitching, or welding.
As a result, these pieces exhibit complementary edges and can be seamlessly combined into larger shapes without gaps.
This situation frequently occurs in the so-called \textit{cutting instances}, where pieces are obtained by partitioning a rectangular sheet that exactly matches the bin.

\begin{figure}[ht!]
    \centering
    \includegraphics[width=0.8\linewidth]{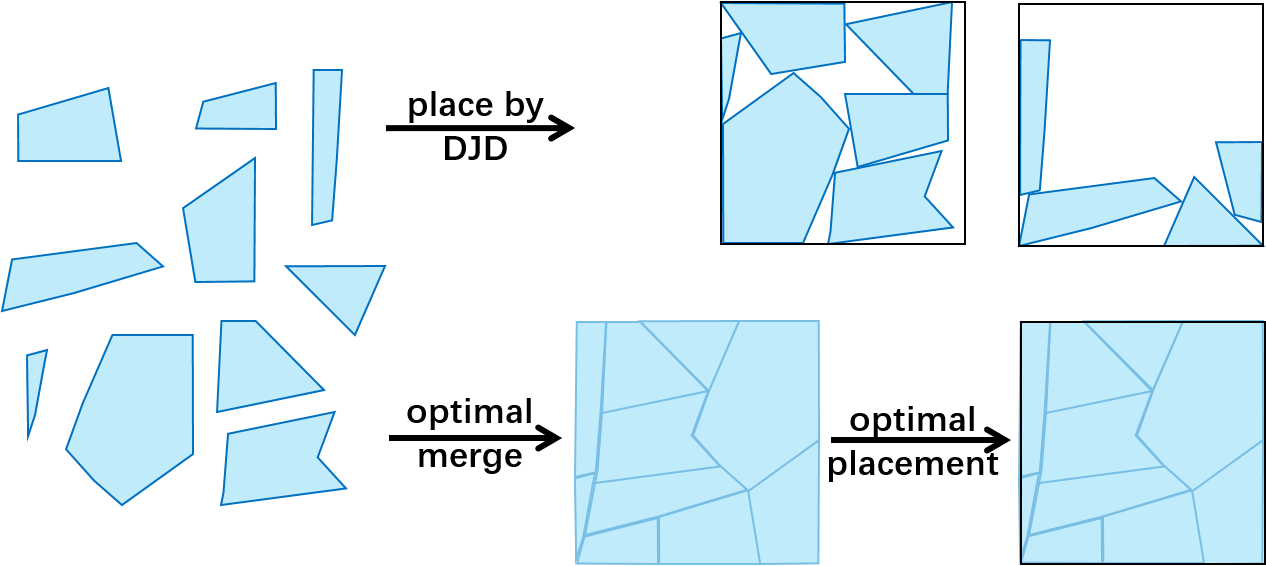}
    \caption{All pieces in a cutting instance can be merged into a single rectangle. However, DJD does not identify this structure, resulting poor solution.}
    \label{fig:mergenew}
\end{figure}

For example, Figure~\ref{fig:mergenew} illustrates a cutting instance in which pieces can be perfectly combined into a single rectangle.
However, DJD does not identify this structure because adjacency is determined solely based on bin utilization during placement.

\begin{figure}[ht!]
    \centering
    \includegraphics[width=1\linewidth]{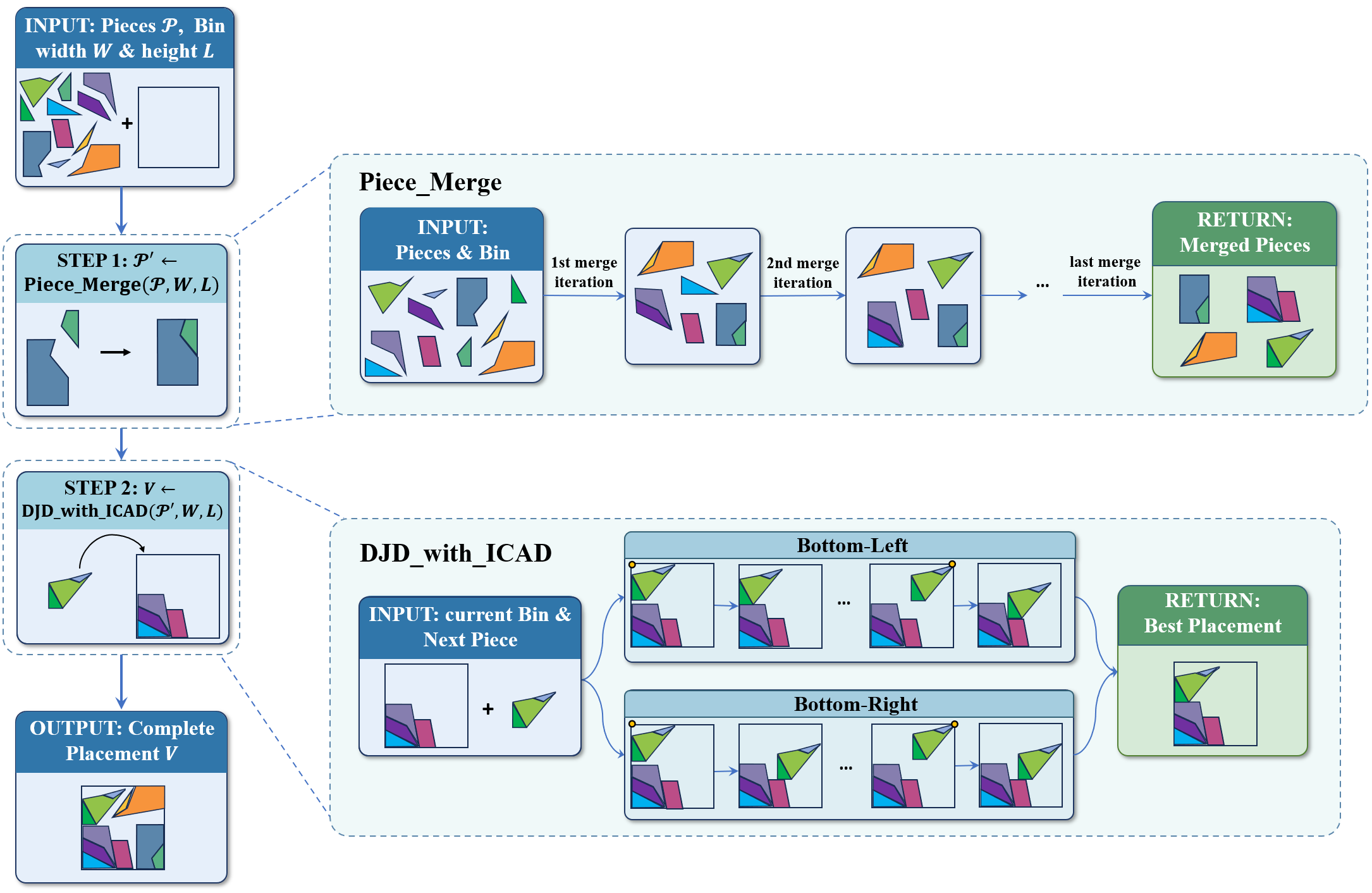}
    \caption{The overall process of the proposed MergeDJD algorithm. The Piece\_Merge procedure is detailed in  Algorithm \ref{alg:groupmerge}; The DJD\_with\_ICAD procedure is detailed in Algorithm \ref{alg:DJD} and Algorithm \ref{alg:ICAD}.}
    \label{fig:all}
\end{figure}

To address this limitation, we introduce a preprocessing algorithm over the DJD, namely the \emph{piece merge algorithm}.
The piece merge algorithm tries to identify the pieces that should be placed together, followed by the piece placement algorithm over the merged pieces.
Therefore, the overall algorithm, namely MergeDJD, as sketched in Figure~\ref{fig:all}, consists of two main components: (i) piece merge and (ii) piece placement.
As mentioned, the piece merge component iteratively tests a combination of pieces, determines whether it can form a large shape, and, if affirmative, replaces the original individual pieces.
The piece placement component determines a complete and feasible placement for the merged piece set. Specifically, it uses DJD to determine the bin assignment and placement order. However, the final placement positions are computed by an improved CAD heuristic (ICAD).
ICAD extends the original CAD by incorporating a bottom-right strategy and additional initial placement points. It is more effective placement regarding non-convex and merged pieces.



\section{The Piece Merge Algorithm}
\label{sec:merge} 
Given a set of pieces, there exist multiple ways to combine them into larger composite pieces.
Even when only two pieces are considered, different geometric configurations can lead to different combined results.
In this section, we investigate the \emph{piece merge algorithm} which limits the way of combining two pieces but keeps the effectiveness.

Intuitively, the piece merge algorithm greedily attempts to combine pairs of pieces by overlapping their corner points.
Once two pieces are successfully combined, the resulting \emph{combined piece} replaces the original pieces and is treated as a single entity in subsequent iterations.
The combined piece is represented by the union of the point sequences of the original pieces.
The process enables multiple rounds of pairwise combination and dynamically updates the input set of pieces.
For cutting instances derived from a rectangular polygon, the algorithm can, in principle, recover a rectangular shape through successive combinations.
Note that, when the relative position of pieces in a combined piece is fixed in the sequential procedure.

\subsection{The Details of Piece Merge Algorithm}

We show the pseudo-code for the piece merge algorithm in Algorithm~\ref{alg:groupmerge}.
Specifically, the algorithm maintains a working set of pieces $\mathcal{P}'$, which is initially equal to the input set of pieces $\mathcal{P}$.
It repeatedly attempts to combine pairs of pieces as long as at least one acceptable combined piece exists (line~\ref{line:groupmerge while}).
At the beginning of each iteration, the pieces in $\mathcal{P}'$ are sorted in descending order of area.
The algorithm then enumerates all unordered pairs $(P_i,P_j)$, where $P_i$ precedes $P_j$ in the sorted list.

\begin{figure}[ht!]
    \centering
    \includegraphics[width=0.8\linewidth]{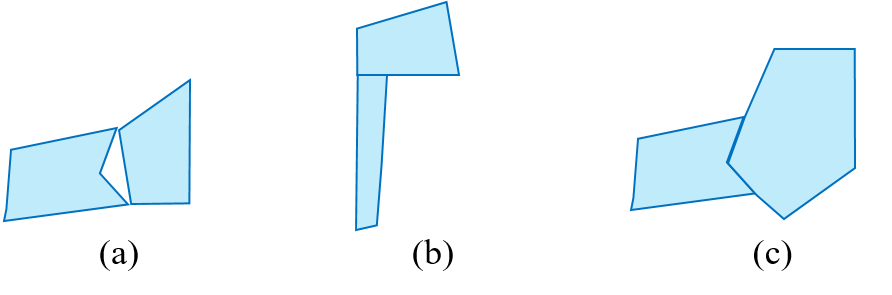}
    \caption{Examples of pieces merged from two pieces. (a) The two pieces touch at two points.  (b) The two pieces overlap along a segment. (c) The two pieces fit perfectly at the corner.}
    \label{fig:merge_illustration}
\end{figure}

From line~\ref{line:exhaustively check begin} to line~\ref{line:exhaustively check end}, the algorithm exhaustively checks whether the two pieces $P_i$ and $P_j$ can be combined under the given bin size and threshold $T$.
Given a pair $(P_i,P_j)$, the algorithm enumerates all allowed rotations $\alpha\in\mathcal{A}$ of $P_j$.
For each rotation, all vertex points $(x_i,y_i)$ of $P_i$ and $(x_j,y_j)$ of $P_j$ are considered.
The piece $P_j$ is translated so that the two selected vertices coincide, giving a candidate combined piece $P_k$.
If $P_i$ and $P_j$ do not overlap and the bounding box of $P_k$ fits within an empty bin, then $P_k$ is considered a feasible combined piece.
Examples of feasible combined pieces are shown in Figure~\ref{fig:merge_illustration}.

When a feasible combined piece is obtained, its fitness value is further evaluated.
If the fitness value is no less than a threshold $T$, the combined piece is accepted and the enumeration is terminated.
The original pieces $P_i$ and $P_j$ are removed from $\mathcal{P}'$, and the combined piece $P_k$ is inserted into the set.
The algorithm then continues to try other combinations on the updated working set.

The procedure ends when no pair of pieces in $\mathcal{P}'$ can be further combined (line~\ref{line:groupmerge while}).
Since each successful combination reduces the number of elements in $\mathcal{P}'$ by one, the number of merge attempts is bounded by the number of input pieces $|\mathcal{P}|$.
The final set of combined pieces $\mathcal{P}'$ is then returned.



\begin{algorithm} [ht!]
\caption{The Piece Merge algorithm}
\label{alg:groupmerge}
\textbf{Input:} a set of pieces $\mathcal{P}$, the width $W$ and the height $L$ of each bin, the threshold parameter $T\in \mathbb{R}$. \\
\textbf{Output:} a new set of pieces $\mathcal{P}'$ after merging.
\begin{algorithmic}[1]
    \State initialize $\mathcal{P}'\leftarrow \mathcal{P}$, $cnt\leftarrow 0$
    \While{$|\mathcal{P}'|!=cnt$} \label{line:groupmerge while}
    \State sort pieces in $\mathcal{P}'$ in descending order of area, $cnt\leftarrow |\mathcal{P}'|$
    \For{$i=1,\ldots ,|\mathcal{P}'|-1$}
    \For{$j=|\mathcal{P}'|,\ldots , i+1$}
    \For{each rotation $\alpha \in \mathcal{A}$} \label{line:exhaustively check begin} \Comment{exhaustively check whether the two pieces can be merged}
        \State rotate $P_j$ with $\alpha$ degrees in clockwise direction
        \For{each point $(x_i,y_i)$ of $P_i$}
            \For{each point $(x_j,y_j)$ of $P_j$}
                \State $d_x\gets x_j-x_i$, $d_y\gets y_j-y_i$
                \State fix $P_i$ and move $P_j$ along vector $(d_x, d_y)$ to get $P_k$
                \If{$P_i$ and $P_j$ do not overlap, $\overline{x_{P_k}}-\underline{x_{P_k}}\leq W$,  $\overline{y_{P_k}}-\underline{y_{P_k}}\leq L$ and $f(P_i,P_j,P_k)\geq T$ } \Comment{equation~\eqref{eq:f}} 
                    \State  $\mathcal{P}' \gets$ $(\mathcal{P}'\cup$ $\{ P_k \})\setminus \{ P_i,P_j \}$, \textbf{break}
                \EndIf
            \EndFor
            \If{$\{ P_i,P_j \}\nsubseteq \mathcal{P}'$} \textbf{break} \EndIf
        \EndFor
        \If{$\{ P_i,P_j \}\nsubseteq \mathcal{P}'$} \textbf{break} \EndIf
    \EndFor \label{line:exhaustively check end}
    \EndFor
    \EndFor
    \EndWhile  \label{line:groupmerge endwhile}
    \State \Return $\mathcal{P}'$
    \end{algorithmic}
\end{algorithm}
    
\subsection{Evaluating the Combined Piece}
\label{sec: evaluate}

Given two pieces $P_1$ and $P_2$, each of which may be a combined piece, let $P_{1,2}$ denote the \emph{combined piece} obtained by merging them.
We define a fitness function to evaluate whether replacing $P_1$ and $P_2$ with $P_{1,2}$ is beneficial for subsequent placement.
The fitness function is defined as
\begin{align}
f(P_1,P_2,P_{1,2}) =
g_{vtx}(P_1,P_2,P_{1,2})
+ g_{peri}(P_1,P_2,P_{1,2})
+ g_{rect}(P_{1,2}).
\label{eq:f}
\end{align}

The function $f$ is an equally weighted sum of three components.
The first two components evaluate the structural simplification achieved by merging $P_1$ and $P_2$, while the third measures the global regularity of the combined piece.
Note that $g_{rect}$ depends only on $P_{1,2}$.



\begin{itemize}
    \item \textit{$g_{vtx}$: vertex number reduction}.  
    Given a polygon $P$, let $|P|$ denote the number of vertices of $P$.
    The vertex reduction term is defined as
    \[
    g_{vtx}(P_1,P_2,P_{1,2}) =
    \max\!\left(0, \max(|P_1|,|P_2|) - |P_{1,2}|\right).
    \]
    When computing the number of points of a combined piece, such as $|P_{1,2}|$, identical vertices shared by $P_1$ and $P_2$ are counted only once.
    Intuitively, a shape described by fewer vertices is simpler.
    Therefore, merges that significantly reduce the number of vertices are favored.
    This criterion is particularly effective when merging non-convex pieces, which often require many vertices to represent concave structures.

    \item \textit{$g_{peri}$: perimeter reduction}.  
   The segments of a polygon are line segments connecting its consecutive points.
    Let $Seg(P)$ denote the set of all boundary segments of the polygons in $P$.
    The length of a segment $e$ is denoted by $len(e)$.
    Given two pieces $P_1$ and $P_2$, the overlapping length of two segments $e_i\in Seg(P_1),e_j\in Seg(P_2)$ in the combined piece $P_{1,2}$ is computed using Algorithm~11 in \citet{López-Camacho2013}, denoted by $overlap\_len(P_{1,2},e_i,e_j)$.
    The perimeter reduction term is defined as
    \[
    g_{peri}(P_1,P_2,P_{1,2}) =
    \sum_{e_i\in Seg(P_1)} \sum_{e_j\in Seg(P_2)}
    \left(
    \frac{overlap\_len(P_{1,2},e_i,e_j)}
         {\max(len(e_i),len(e_j))}
    \right)^2.
    \]
    This term measures the extent to which the perimeters of $P_1$ and $P_2$ overlap in the combined piece.
    Since shapes with shorter perimeters are typically simpler, merges that reduce the overall perimeter are preferred.

    \item \textit{$g_{rect}$: global regularity of $P_{1,2}$}.  
    Let $\overline{x}$ and $\underline{x}$ denote the maximum and minimum $x$-coordinates of the vertices of $P_{1,2}$, and define $\overline{y}$ and $\underline{y}$ analogously.
    The envelope rectangle of $P_{1,2}$ is the axis-aligned rectangle defined by these extrema.
    The regularity term is defined as
    \[
    g_{rect}(P_{1,2}) =
    \left(
    \frac{Area(P_{1,2})}
         {(\overline{x}-\underline{x})(\overline{y}-\underline{y})}
    \right)^2.
    \]
    This term evaluates how closely the combined piece resembles a rectangle.
    Intuitively, shapes that are closer to rectangular are more likely to be placed efficiently in subsequent packing stages.
\end{itemize}

\section{Improving the CAD Placement Strategy}
\label{CAD}


As mentioned, the placement heuristic, namely the CAD algorithm, plays an important role in the DJD algorithm \citep{López-Camacho2013}.
The CAD algorithm was originally proposed by \citet{uday2001nesting,hifi2003hybrid}. 
In CAD, there are mainly two crucial steps, setting the initial positions and the placement strategy.

\paragraph{Setting the Initial Positions}
Given the a piece $P$ with rotation $\alpha$ to be placed in a bin $i$, CAD constructs a set of two-dimensional points, denoted as $cand$, for the potential placement positions of the reference point of the piece. 
Firstly, CAD puts the four corners of the bin $(0,0),(0,W),(0,L),(W,L)$ into $cand$.
For each piece $P'$ that has been placed in the bin, CAD introduces five points ($\bar{x}, 0$), ($0, \bar{y}$), ($\underline{x}, \bar{y}$), ($\bar{x}, \bar{y}$), ($\bar{x}, \underline{y}$) where $\bar{x}$ (or $\underline{x}$), $\bar{y}$ (or $\underline{y}$) are the maximum (or minimum) coordinates in x-coordinates and y-coordinates, respectively. 
\begin{figure}[ht!]
    \centering
    \includegraphics[width=0.8\linewidth]{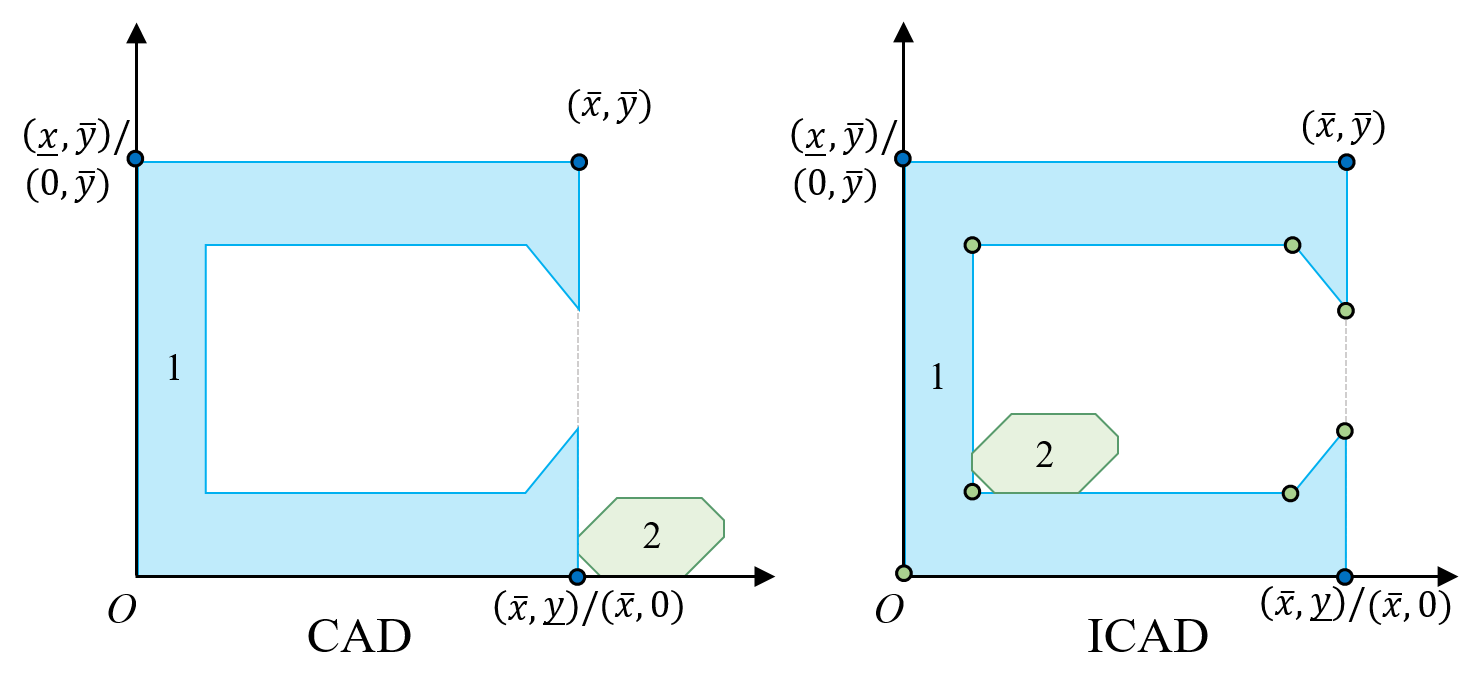}
    \caption{An illustration of the effect of initial positions. Left: the placement generated by CAD.  After placing piece 1, the initial positions for piece 2 generated by CAD are colored in blue. Right: the placement generated by ICAD. After placing piece 1, the augmented initial positions for piece 2 generated by ICAD are colored in green. By introducing additional initial positions, ICAD enables placements inside the concave region, resulting in a higher utilization ratio.}
    \label{fig:add}
\end{figure}
\begin{figure}[ht!]
    \centering
    \includegraphics[width=0.6\linewidth]{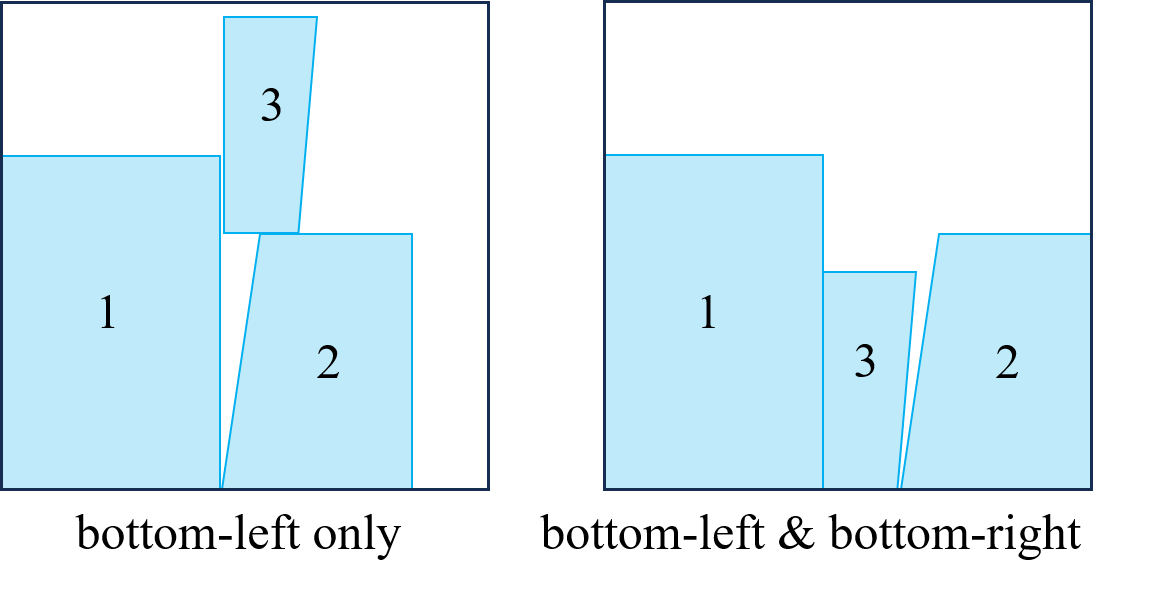}
    \caption{Placements generated by the bottom-left-only strategy (used in CAD) and by the mixed bottom-left \& bottom-right strategy (used in ICAD).
In the mixed strategy, the bottom-left strategy is applied to Pieces 1 and 3, while the bottom-right strategy is applied to Piece 2, which leads to a more compact placement.}
    \label{fig:bottomrightcompare}
\end{figure}

\paragraph{Placement Strategy} For each initial position $(x_0,y_0)\in cand$, CAD firstly moves the reference point of the piece $P$ to this position. After this move, if no two pieces share overlapping area and the coordinates of $P$ are within the bin, then this merge is legal. Next, CAD iteratively slides the piece $P$ downwards and then leftwards until it cannot be slide further.
Without loss of generality, assume this position is $(x_t,y_t)$. During the whole process, the piece rotation is fixed as $\alpha$.
This move generates a position triple of $P$, indicated by $(i,\vec{v}=(x_t,y_t),\alpha)$.
The $g_{peri}$ metric of the shape of union pieces after placing $P$ is calculated according to the position triple.
The CAD enumerates all initial positions for $P$ and return the position triple that has highest $g_{peri}$.

\begin{algorithm} [ht!]
\caption{The ICAD algorithm}
\label{alg:ICAD}
\textbf{Input:} a bin $i$, and a group of ordered pieces $\vec{O}$. 
\\
\textbf{Output:} success with placement positions, or failure.
\begin{algorithmic}[1]
\State $cand\leftarrow \{(0,0),(0,L),(W,0),(W,L)\}$
\For{each placed piece $P$ in bin $i$}
\State $cand\leftarrow cand\cup \{(0,\overline{y_P}),(\underline{x_P},\overline{y_P}),(\overline{x_P},\overline{y_P}),(\overline{x_P},\underline{y_P}),(\overline{x_P},0)\}\cup P$
\EndFor
\State denote the current bin with placed pieces as a combined piece $P_B$
\For{$j=1,\ldots,|\vec{O}|$}
    \State $cur\_best\leftarrow -1$
    \For{$\alpha \in \mathcal{A}$}
        \State rotate piece $P_j$ in counterclockwise by $\alpha$ degree around its reference point.
        \For{each initial position $(x,y)\in cand$}
        \If{after placing $P_j$'s reference point at $(x,y)$, no two pieces overlap, $\underline{x_{P_j}},\overline{x_{P_j}}\in [0,W]$ and  $\underline{y_{P_j}},\overline{y_{P_j}}\in [0,L]$ }
        \State $v_{P_j,L}\leftarrow bottom\_left\_strategy(i,P_j,(x,y))$
        \State denote the bin after applying $v_{P_j,L}$ as a combined piece $P_{BL}$
        \If{$g_{peri}(P_B,P_j,P_{BL})> cur\_best$}
        \State $v_{P_j}\leftarrow v_{P_j,L}$, $cur\_best\leftarrow g_{peri}(P_B,P_j,P_{BL})$
        \EndIf
        \State $v_{P_j,R}\leftarrow bottom\_right\_strategy(i,P_j,(x,y))$
        \State denote the bin after applying $v_{P_j,R}$ as a combined piece $P_{BR}$
        \If{$g_{peri}(P_B,P_j,P_{BR})> cur\_best$}
        \State $v_{P_j}\leftarrow v_{P_j,R}$, $cur\_best\leftarrow g_{peri}(P_B,P_j,P_{BR})$
        \EndIf
        \EndIf
        \EndFor
        \If{$cur\_best=-1$} \Return failure
        \EndIf
    \EndFor
\EndFor
\State \Return success,$\{v_{P}\mid \forall P\in \vec{O}\}$
\end{algorithmic}
\end{algorithm}

\subsection{Improve CAD to ICAD}
We improve the CAD algorithm by increasing the set of initial placement positions and introducing an additional placement strategy. The pseudo-code of the improved algorithm is given in Algorithm~\ref{alg:ICAD}. 
In terms of initial positions, we further increase $cand$ with the points of the pieces placed. 
As shown in Figure \ref{fig:add}, the $cand$ used in CAD cannot place the piece inside the large concave region, leading to an inferior utilization ratio. 
In contrast, the increased $cand$ enables the use of this large concave region, which may significantly improve the utilization ratio.

In terms of placement strategies, we improve CAD with the bottom-right move, that is, slide the piece $P$ downward and then rightward until it cannot slide further. 
This move generates a position triple that represents a placement of $P$. 
The placed piece $P$ is treated as one piece, while the bin together with all previously placed pieces is treated as another piece. These two pieces form a combined piece.
The combined piece is then evaluated using the $g_{peri}$ metric.
As shown in Figure \ref{fig:bottomrightcompare}, the incorporation of bottom-right move enables the generation of a more compact placement.

\section{Computational Experiments}
\label{result}

\subsection{Datasets}
\label{instances}

\begin{table}[ht!]
\centering
 \caption{ Statistics of JP1 and JP2 instances. All instances share a bin size $1000\times 1000$.}
 \label{tab:JP1 and JP2}
\resizebox{\linewidth}{!}{
\begin{tabular}{|cccc||cccc|}
\hline
JP1 & \#pieces & \#instances & opt \#bins & JP2 & \#pieces & \#instances & opt \#bins \\ \hline
Conv A             & 30           & 30               & 3                 & NConv A             & 35-50           & 30               & 3                 \\ 
Conv B             & 30           & 30               & 10                & NConv B             & 40-52           & 30               & 10                \\ 
Conv C             & 36           & 30               & 6                 & NConv C             & 42-60           & 30               & 6                 \\ 
Conv D             & 60           & 30               & 3                 & NConv F             & 35-45           & 30               & 2                 \\ 
Conv E             & 60           & 30               & 3                 & NConv H             & 42-60           & 30               & 12                 \\ 
Conv F             & 30           & 30               & 2                 & NConv L             & 35-45           & 30               & 3                 \\ 
Conv G             & 36           & 30               & $\leq$15          & NConv M             & 45-58           & 30               & 5          \\ 
Conv H             & 36           & 30               & 12                & NConv O             & 33-43           & 30               & 7                \\ 
Conv I             & 60           & 30               & 3                 & NConv S             & 17-20           & 30               & 2                 \\ 
Conv J             & 60           & 30               & 4                 & NConv T             & 30-40           & 30               & 10                 \\ 
Conv K             & 54           & 30               & 6                 & NConv U             & 20-33           & 30               & 5                 \\ 
Conv L             & 30           & 30               & 3                 & NConv V             & 15-18           & 30               & 5                 \\ 
Conv M             & 40           & 30               & 5                 & NConv W             & 24-28           & 30               & 4                 \\ 
Conv N             & 60           & 30               & 2                 & NConv X             & 25-39           & 30               & 3                 \\ 
Conv O             & 28           & 30               & 7                 & NConv Y             & 40-50           & 30               & 6                 \\ 
Conv P             & 56           & 30               & 8                 & NConv Z             & 60           & 30               & 12                 \\ 
Conv Q             & 60           & 30               & 15                &              &            &                &                 \\ 
Conv R             & 54           & 30               & 9                 &              &            &                &                  \\ \hline
Total            &            & 540               &                  &             &            & 480               &                  \\ \hline
\end{tabular}
}
\end{table}

We use two standard datasets the \textit{jigsaw puzzles (JP)} instances and \textit{nesting} instances as in the existing work \citet{zhang2022iteratively}.
These datasets are available on the ESICUP website \url{https://github.com/ESICUP/datasets}.
\begin{itemize}
    \item {\em JP}: 
The JP dataset (also called Terashima dataset) includes two sets of instances JP1 and JP2, proposed by \citet{López-Camacho2013} and \citet{lopez2014unified} respectively. The pieces in these instances are cut from several bins, implying that the optimal solution has an $F$ value of 100$\%$. 
The JP1 dataset contains 18 categories, each comprising 30 instances, for a total of 540 instances. All pieces in JP1 are convex polygons.
The JP2 dataset contains 16 categories, each with 30 instances, for a total of 480 instances. In contrast to JP1, the JP2 dataset includes both convex and non-convex pieces. The statistics of JP instances are shown in Table~\ref{tab:JP1 and JP2}.
    \item {\em Nesting:}
The nesting dataset are derived from 23 distinct instances of strip packing problems where the length of a bin is assumed infinite.
The nesting dataset contains both convex and non-convex polygons.
To adapt these instances in the bin packing problem, we follow the setting of \citet{Martinez-Sykora2017} and define three sizes of bins.
Let $d_{max}$ be the maximum length or width of all pieces in their initial orientation in the given instance. The three container dimensions are set as follows: the small bin (\textbf{SB}) has both its length and width set to $1.1d_{max}$, the medium bin (\textbf{MB}) to $1.5d_{max}$ and the large bin (\textbf{LB}) to $2d_{max}$. The statistics of nesting instances are shown in Table~\ref{tab:nesting}.
\end{itemize}

\begin{table}[ht]
\caption{Statistics of nesting instances}
\label{tab:nesting}
\resizebox{\linewidth}{!}{
\begin{tabular}{|cc|ccc||cc|ccc|}
\hline
\multirow{2}{*}{instance} & \multirow{2}{*}{\#pieces} & \multicolumn{3}{c||}{bin size $W=L$}                              & \multirow{2}{*}{instance} & \multirow{2}{*}{\#pieces} & \multicolumn{3}{c|}{bin size $W=L$}                                \\ \cline{3-5} \cline{8-10} 
                          &                           & {SB}     & {MB}     & LB   &                           &                           & {SB}     & {MB}     & LB     \\ \hline
albano                    & 24                        & {3337.4} & {4551}   & 6068 & poly3b                    & 45                        & {14.3}   & {19.5}   & 26     \\ 
dighe1                    & 16                        & {72.6}   & {99}     & 132  & poly4a                    & 60                        & {14.3}   & {19.5}   & 26     \\ 
dighe2                    & 10                        & {77}     & {105}    & 140  & poly4b                    & 60                        & {14.3}   & {19.5}   & 26     \\ 
fu                        & 12                        & {15.4}   & {21}     & 28   & poly5a                    & 75                        & {14.3}   & {19.5}   & 26     \\ 
han                       & 23                        & {25.3}   & {34.5}   & 46   & poly5b                    & 75                        & {14.3}   & {19.5}   & 26     \\ 
jakobs1                   & 25                        & {8.8}    & {12}     & 16   & shapes0                   & 43                        & {15.4}   & {21}     & 28     \\ 
jakobs2                   & 25                        & {17.6}   & {24}     & 32   & shapes1                   & 43                        & {15.4}   & {21}     & 28     \\ 
mao                       & 20                        & {1206.7} & {1645.5} & 2194 & shapes2                   & 28                        & {5.5}    & {7.5}    & 10     \\ 
poly1a                    & 15                        & {14.3}   & {19.5}   & 26   & shirts                    & 99                        & {26}     & {19.5}   & 14.3   \\ 
poly2a                    & 30                        & {14.3}   & {19.5}   & 26   & swim                      & 48                        & {2133.6} & {2909.5} & 3879.3 \\ 
poly2b                    & 30                        & {14.3}   & {19.5}   & 26   & trousers                  & 64                        & {64.9}   & {88.5}   & 118    \\ 
poly3a                    & 45                        & {14.3}   & {19.5}   & 26   &                           &                           & {}       & {}       &        \\ \hline
\end{tabular}
}
\end{table}

\subsection{Benchmark Algorithms}
To evaluate the effectiveness of MergeDJD, we compare it with the original DJD algorithm\citep{López-Camacho2013}. 
Both DJD and MergeDJD are implemented in Java.
All experiments are conducted on a computer equipped with an Intel(R) Core(TM) i7-11800H @ 2.30GHz processor and 16 GB of RAM.

During the collection of best known values (BKV), we observed that the calculation of the $F$ metric is not always consistent across existing studies. 
Although $F$ is commonly defined as the average of squared utilization ratios of bins \citep{lopez2013understanding,zhang2022iteratively}, as in Section \ref{subsec_metric_definition}, some existing work appears to report values computed using alternative formulations, such as averaging the utilization ratios without squaring. 
As a result, the values $F$ reported  in the literature are not always directly comparable. 
Similar discrepancies can be observed in the results reported for several nesting instances in \citet{yao2024iteratively,luo2025decimal}. 
However, without source codes for these algorithms or data for their final results, it is inappropriate to directly compare $F$ reported in these works. 
Therefore, when reporting the best known values of $F$ and $K$, we do not include the results in \citet{yao2024iteratively,luo2025decimal}, although this potentially overlooks some better solutions. (Indeed, this further highlights the need for open-source implementations and unified benchmark data for the problem.)
In addition, we consistently follow the definition of the $F$ metric introduced in Section \ref{subsec_metric_definition} throughout all experiments.

\subsection{Performance Comparison}
We compare the overall performance of benchmark algorithms in JP1, JP2, nesting-SB, nesting-MB and nesting-LB instances. The results are shown in Tables~\ref{tab:JP1result},\ref{tab:jp2result},\ref{tab:nestingresult-SB},\ref{tab:nestingresult-MB} and \ref{tab:nestingresult-LB} respectively.

\begin{table}[h!]
  \caption{Comparisons on the JP1 instances. Bold values indicate the best results among the compared algorithms. Underlined values denote results that meet or outperform the best known values (BKV). For convenience, the sources of the BKV are labeled as $a$: \citet{zhang2022iteratively}, $b$: \citet{Martinez-Sykora2017}, $c$: \citet{cai2023heuristics}, $d$: \citet{abeysooriya2018jostle}, $e$: \citet{Liu2020}, $f$: \citet{Guerriero}, $g$: \citet{López-Camacho2013}.}
  \label{tab:JP1result}
  \centering
\resizebox{0.8\linewidth}{!}{
\begin{tabular}{|cc|cc|ccc|cllcc|}
\hline
\multirow{3}{*}{\begin{tabular}[c]{@{}c@{}}instance\\ set\end{tabular}} & \multirow{3}{*}{\begin{tabular}[c]{@{}c@{}}ave\\ \#pieces\end{tabular}} & \multicolumn{2}{c|}{BKV} & \multicolumn{3}{c|}{DJD}                & \multicolumn{5}{c|}{MergeDJD}                                                                                                                                                                 \\ \cline{3-12} 
                                                                        &                                                                         & F          & K           & F              & K      & time          & \begin{tabular}[c]{@{}c@{}}ave \#pieces \\ after merge\end{tabular} & \multicolumn{1}{c}{F} & \multicolumn{1}{c}{K} & time          & \begin{tabular}[c]{@{}c@{}}merge\\ time\end{tabular} \\
                                                                 
                                                                   \hline
Conv A                                                                   &30                                                                     & 0.869$^f$      & 3.435$^a$       & 0.586          & 4.000  & 0.62          & 14.2                                                                & {\textbf{0.763}}                 & \underline{\textbf{3.371}}                & \textbf{0.30} & 0.05                                                 
                                            \\
Conv B                                                                  & 30                                                                      & 1.000$^a$      & 10.000$^a$      & 0.899          & 10.900 & 0.51          & 20.0                                                                & \underline{\textbf{1.000}}                & \underline{\textbf{10.000}}                & \textbf{0.14} & 0.01                                                 \\
Conv C                                                                  & 36                                                                      & 0.970$^f$     & 6.580$^a$       & 0.739          & 7.198  & 0.57          & 14.6                                                                & \textbf{0.917}                 & \underline{\textbf{6.207}}                 & \textbf{0.24} & 0.07                                                 \\
Conv D                                                                  & 60                                                                      & 0.610$^d$      & 3.596$^a$       & 0.428          & 5.000  & \textbf{0.66} & 30.7                                                                & \textbf{0.601}                & \textbf{3.879}                 & 1.98          & 0.41                                                 \\
Conv E                                                                  & 60                                                                      & 0.579$^a$      & 3.836$^a$       & 0.327          & 5.667  & \textbf{0.63} & 30.8                                                                & \textbf{0.503}                 & \textbf{4.285}                 & 2.18          & 0.46                                                 \\
Conv F                                                                  & 30                                                                      & 0.648$^f$      & 2.381$^a$       & 0.460          & 3.067  & 0.56          & 12.4                                                                & \textbf{0.616}                & \textbf{2.435}                 & \textbf{0.31} & 0.07                                                 \\
Conv G                                                                  & 36                                                                      & 0.822$^a$      & 13.131$^a$      & 0.814          & 13.458 & 0.50          & 26.6                                                                & \underline{\textbf{0.828}}                 & \underline{\textbf{13.118} }              & \textbf{0.25} & 0.05                                                 \\
Conv H                                                                  & 36                                                                      & 1.000$^a$      & 12.000$^a$      & 0.908          & 12.965 & 0.50          & 24.0                                                                & \underline{\textbf{1.000}}                & \underline{\textbf{12.000}}                & \textbf{0.16} & 0.01                                                 \\
Conv I                                                                  & 60                                                                      & 0.697$^d$      & 3.250$^a$      & 0.621          & 4.000  & \textbf{0.59} & 60.0                                                                & \textbf{0.639}                & \textbf{3.810}                & 8.96          & 0.01                                                 \\
Conv J                                                                  & 60                                                                      & 0.701$^b$      & 4.467$^a$       & 0.639          & 5.100  & \textbf{0.59} & 40.7                                                                & \textbf{0.680}                 & \textbf{4.875}                 & 1.92          & 0.03                                                 \\
Conv K                                                                  & 54                                                                      & 0.923$^f$      & 6.641$^a$       & 0.664          & 7.800  & 0.58          & 33.8                                                                & \textbf{0.840}                & \underline{\textbf{6.460}}                 & \textbf{0.47} & 0.15                                                 \\
Conv L                                                                  & 30                                                                      & 0.630$^d$      & 3.667$^a$       & 0.438          & 4.894  & 0.55          & 13.2                                                                & \underline{\textbf{0.716 } }               & \underline{\textbf{3.492}}                & \textbf{0.50} & 0.14                                                 \\
Conv M                                                                  & 40                                                                      & 0.723$^d$      & 5.936$^a$       & 0.540          & 7.155  & 0.62          & 16.6                                                                & \underline{\textbf{0.794}}                 & \underline{\textbf{5.582}}                 & \textbf{0.52} & 0.17                                                 \\
Conv N                                                                  & 60                                                                      & 0.688$^b$      & 2.360$^a$       & 0.467          & 3.000  & \textbf{0.82} & 30.0                                                                & \textbf{0.519}                 & \textbf{2.793}                 & 2.20          & 0.26                                                 \\
Conv O                                                                  & 28                                                                      & 0.994$^f$      & 7.005$^a$       & 0.844          & 7.878  & 0.51          & 8.7                                                                 & \textbf{0.898}                 & \textbf{7.345}                 & \textbf{0.32} & 0.05                                                 \\
Conv P                                                                  & 56                                                                      & 0.852$^b$      & 9.190$^a$       & 0.632          & 10.610 & \textbf{0.66} & 28.7                                                                & \textbf{0.768}                & \textbf{9.243}                & 1.34          & 0.47                                                 \\
Conv Q                                                                  & 60                                                                      & 1.000$^a$      & 15.000$^a$      & \underline{\textbf{1.000}}          & \underline{\textbf{15.000}} & 0.49          & 30.2                                                                & \underline{\textbf{1.000}}                 & \underline{\textbf{15.000}}               & \textbf{0.26} & 0.01                                                 \\
Conv R                                                                  & 54                                                                      & 0.946$^f$      & 10.064$^a$      & 0.716          & 10.963 & 0.77          & 26.0                                                                & \textbf{0.872}                 & \underline{\textbf{9.573}}                 & \textbf{0.52} & 0.20                                                 \\ \hline
\multicolumn{2}{|c|}{average}                                                                                                                     & 0.814      & 6.808       & 0.651          & 7.703  & \textbf{0.60} & 25.6                                                                & \textbf{0.775}                 & \textbf{6.859}                 & 1.26          & 0.14                                                 \\
\multicolumn{2}{|c|}{min}                                                                                                                         & 0.579      & 2.360       & 0.327          & 3.000  & 0.49          & 8.7                                                                 & \textbf{0.503}                 & \textbf{2.435}                 & \textbf{0.14} & 0.01                                                 \\
\multicolumn{2}{|c|}{Q1}                                                                                                                          & 0.690      & 3.614       & 0.485          & 4.921  & 0.52          & 15.1                                                                & \textbf{0.649}                & \textbf{3.827}                 & \textbf{0.27} & 0.03                                                 \\
\multicolumn{2}{|c|}{Q2}                                                                                                                          & 0.837      & 6.258       & 0.636          & 7.176  & 0.58          & 26.3                                                                & \textbf{0.781}                 & \underline{\textbf{5.894}}                 & \textbf{0.48} & 0.07                                                 \\
\multicolumn{2}{|c|}{Q3}                                                                                                                          & 0.964      & 9.798       & 0.795          & 10.827 & \textbf{0.63} & 30.6                                                                & \textbf{0.892}                 & \underline{\textbf{9.491}}                 & 1.78          & 0.19                                                 \\
\multicolumn{2}{|c|}{max}                                                                                                                         & 1.000      & 15.000      & \underline{\textbf{1.000}} & \underline{\textbf{15.000}} & \textbf{0.82} & 60.0                                                                & \underline{\textbf{1.000}}                & \underline{\textbf{15.000}}                & 8.96          & 0.47                                                 \\ \hline
\end{tabular}
}
\end{table}

\begin{table}[h!]
  \centering
  \caption{Comparisons on the JP2 instances. Bold values indicate the best results among the compared algorithms. Underlined values denote results that meet or outperform the best known values (BKV).}
  \label{tab:jp2result}
\resizebox{0.8\linewidth}{!}{
\begin{tabular}{|cc|cc|ccc|cllcc|}
\hline
\multirow{3}{*}{\begin{tabular}[c]{@{}c@{}}instance\\ set\end{tabular}} & \multirow{3}{*}{\begin{tabular}[c]{@{}c@{}}ave\\ \#pieces\end{tabular}} & \multicolumn{2}{c|}{BKV} & \multicolumn{3}{c|}{DJD} & \multicolumn{5}{c|}{MergeDJD}                                                                                                                                                        \\ \cline{3-12} 
                                                                        &                                                                         & F          & K           & F      & K       & time  & \begin{tabular}[c]{@{}c@{}}ave \#pieces \\ after merge\end{tabular} & \multicolumn{1}{c}{F} & \multicolumn{1}{c}{K} & time & \begin{tabular}[c]{@{}c@{}}merge\\ time\end{tabular} \\
\hline
Nconv A                                                                 & 41.7                                                                    & 0.680$^b$      & 3.527$^a$       & 0.507  & 4.500   & 0.64  & 14.2                                                                & \underline{\textbf{0.680}}                & \underline{\textbf{3.527}}                 & \textbf{0.44} & 0.08                                                 \\
Nconv B                                                                 & 47.3                                                                    & 0.841$^a$      & 10.808$^a$      & 0.745  & 11.981  & 0.84  & 17.9                                                                & \textbf{0.746}                & \textbf{11.646}                & \textbf{0.27} & 0.04                                                 \\
Nconv C                                                                 & 50.0                                                                    & 0.702$^a$      & 7.007$^a$       & 0.612  & 8.099   & 0.74  & 14.8                                                                & \underline{\textbf{0.880}}                 & \underline{\textbf{6.311}}                 & \textbf{0.36} & 0.11                                                 \\
Nconv F                                                                 & 40.0                                                                    & 0.551$^b$      & 2.413$^a$       & 0.437  & 3.200   & 0.71  & 12.6                                                                & \underline{\textbf{0.578}}                 & \textbf{2.511}                 & \textbf{0.49} & 0.11                                                 \\
Nconv H                                                                 & 50.0                                                                    & 0.895$^a$      & 12.654$^a$      & 0.802  & 13.771  & 0.81  & 22.1                                                                & \textbf{0.828}                 & \textbf{13.315}                & \textbf{0.27} & 0.04                                                 \\
Nconv L                                                                 & 40.0                                                                    & 0.591$^a$      & 3.771$^a$       & 0.404  & 5.064   & 0.67  & 13.5                                                                & \underline{\textbf{0.718}}                 & \underline{\textbf{3.495}}                 & \textbf{0.48} & 0.17                                                 \\
Nconv M                                                                 & 51.0                                                                    & 0.681$^b$      & 6.073$^a$       & 0.492  & 7.597   & 0.80  & 16.4                                                                & \underline{\textbf{0.790}}                 & \underline{\textbf{5.557}}                 & \textbf{0.51} & 0.17                                                 \\
Nconv O                                                                 & 38.0                                                                    & 0.879$^a$      & 7.488$^a$       & 0.734  & 8.556   & 0.64  & 9.1                                                                 & \textbf{0.861}                 & \textbf{7.516}                 & \textbf{0.31} & 0.07                                                 \\
Nconv S                                                                 & 19.0                                                                    & 0.817$^a$      & 2.185$^a$       & 0.427  & 3.425   & 0.54  & 6.2                                                                 & \underline{\textbf{0.842}}                 & \textbf{2.233}                 & \textbf{0.30} & 0.04                                                 \\
Nconv T                                                                 & 34.6                                                                    & 0.997$^a$      & 10.019$^a$      & 0.853  & 11.142  & 0.53  & 10.1                                                                & \textbf{0.995}                 & \underline{\textbf{10.018}}                & \textbf{0.23} & 0.02                                                 \\
Nconv U                                                                 & 26.5                                                                    & 0.894$^a$      & 5.242$^a$       & 0.668  & 6.516   & 0.57  & 5.2                                                                 & \underline{\textbf{0.975}}                 & \underline{\textbf{5.046}}                 & \textbf{0.28} & 0.03                                                 \\
Nconv V                                                                 & 16.5                                                                    & 1.000$^a$      & 5.000$^a$       & 0.898  & 5.465   & 0.50  & 5.2                                                                 & \textbf{0.950}                 & \textbf{5.117}                 & \textbf{0.23} & 0.01                                                 \\
Nconv W                                                                 & 26.0                                                                    & 0.843$^d$      & 4.364$^a$       & 0.605  & 5.465   & 0.53  & 14.6                                                                & \underline{\textbf{0.931}}                 & \underline{\textbf{4.137}}                 & \textbf{0.24} & 0.01                                                 \\
Nconv X                                                                 & 32.0                                                                    & 0.649$^d$      & 3.579$^a$       & 0.477  & 4.666   & 0.59  & 11.5                                                                & \underline{\textbf{0.967}}                 & \underline{\textbf{3.038}}                 & \textbf{0.28} & 0.03                                                 \\
Nconv Y                                                                 & 45.0                                                                    & 0.743$^a$      & 6.870$^a$       & 0.628  & 7.965   & 0.72  & 14.4                                                                & \underline{\textbf{0.853}}                 & \underline{\textbf{6.405}}                 & \textbf{0.31} & 0.07                                                 \\
Nconv Z                                                                 & 60.0                                                                    & 0.893$^b$      & 12.843$^a$      & 0.806  & 13.830  & 0.74  & 20.4                                                                & \underline{\textbf{0.897}}                 & \underline{\textbf{12.655}}                & \textbf{0.49} & 0.20                                                 \\ \hline
\multicolumn{2}{|c|}{average}                                                                                                                     & 0.791      & 6.490       & 0.631  & 7.578   & 0.66  & 13.0                                                                & \underline{\textbf{0.843}}                 & \underline{\textbf{6.408}}                 & \textbf{0.34} & 0.07                                                 \\
\multicolumn{2}{|c|}{min}                                                                                                                         & 0.551      & 2.185       & 0.404  & 3.200   & 0.50  & 5.2                                                                 & \underline{\textbf{0.578}}                 & \textbf{2.233}                 & \textbf{0.23} & 0.01                                                 \\
\multicolumn{2}{|c|}{Q1}                                                                                                                          & 0.681      & 3.723       & 0.488  & 4.964   & 0.57  & 9.9                                                                 & \underline{\textbf{0.779}}                 & \underline{\textbf{3.519}}                 & \textbf{0.27} & 0.03                                                 \\
\multicolumn{2}{|c|}{Q2}                                                                                                                          & 0.829      & 5.658       & 0.620  & 7.057   & 0.66  & 13.9                                                                & \underline{\textbf{0.857}}                 & \underline{\textbf{5.337}}                 & \textbf{0.30} & 0.06                                                 \\
\multicolumn{2}{|c|}{Q3}                                                                                                                          & 0.893      & 8.121       & 0.759  & 9.202   & 0.74  & 15.2                                                                & \underline{\textbf{0.936}}                 & \textbf{8.141}                 & \textbf{0.45} & 0.11                                                 \\
\multicolumn{2}{|c|}{max}                                                                                                                         & 1.000      & 12.843      & 0.898  & 13.830  & 0.84  & 22.1                                                                & \textbf{0.995}                 & \textbf{13.315}                & \textbf{0.51} & 0.20                                                 \\ \hline
\end{tabular}
}
\end{table}

Columns 1-2 in Tables~\ref{tab:JP1result}-\ref{tab:jp2result} respectively give the name of each instance set and the average number of pieces in each set. Columns 3-4 report the BKV results on each instance set, including the $F$ metric (higher is better) and the $K$ metric (lower is better). Columns 5-12 summarize the results obtained by the compared algorithms, including the total runtime of algorithms ($time$). For MergeDJD, we also provide the average number of pieces after merge and merge time.
Finally, the summarized results for each column are presented in the last six rows of the table, including the average, min, max and three quartiles Q1, Q2 and Q3. $Q_1$and $Q_3$ correspond to the 25th and 75th percentiles, respectively, while $Q_2$ denotes the median value. These quartiles characterize the central tendency and dispersion of the experimental outcomes in a robust manner. Note that each nesting dataset consists of a single instance. Therefore, columns 1-2 in Tables~\ref{tab:nestingresult-SB}-\ref{tab:nestingresult-LB} respectively give the name of each instance and the number of pieces in each instance.

\begin{table}[h!]
  \centering
  \caption{Comparisons on the Nesting-SB instances. Bold values indicate the best results among the compared algorithms. Underlined values denote results that meet or outperform the best known values (BKV).}
  \label{tab:nestingresult-SB}
    \resizebox{0.8\linewidth}{!}{
    \begin{tabular}{|cc|ll|llc|ccccc|}
\hline
\multirow{3}{*}{\begin{tabular}[c]{@{}c@{}}instance\\ set\end{tabular}} & \multirow{3}{*}{\begin{tabular}[c]{@{}c@{}}ave\\ \#pieces\end{tabular}} & \multicolumn{2}{c|}{BKV} & \multicolumn{3}{c|}{DJD} & \multicolumn{5}{c|}{MergeDJD}                                                                                                                                                        \\ \cline{3-12} 
                                                                        &                                                                         & F          & K           & F      & K       & time  & \begin{tabular}[c]{@{}c@{}}ave \#pieces \\ after merge\end{tabular} & \multicolumn{1}{c}{F} & \multicolumn{1}{c}{K} & time & \begin{tabular}[c]{@{}c@{}}merge\\ time\end{tabular} \\
\hline
albano                    & 24                        & 0.605$^d$                     & 4.677$^a$                       & 0.588                 & \textbf{4.911}        & \textbf{1.07}  & 24                                                              & \textbf{0.589} & 4.979                                    & 16.72         & 2.43                                                 \\
dighe1                    & 16                        & 0.457$^a$                     & 2.472$^a$                       & 0.276                 & 3.873                 & \textbf{0.76}  & 15                                                              & \textbf{0.281} & \textbf{3.386}                           & 1.30          & 0.36                                                 \\
dighe2                    & 10                        & 0.397$^a$                     & 2.390$^a$                       & \textbf{0.367}        & 3.000                 & \textbf{0.74}  & 9                                                               & 0.362          & \textbf{2.870}                           & 0.75          & 0.20                                                 \\
fu                        & 12                        & 0.448$^d$                     & 7.455$^a$                       & \textbf{0.362}        & 7.649                 & 0.48           & 12                                                              & \textbf{0.362} & \textbf{\underline{7.455}}  & \textbf{0.38} & 0.08                                                 \\
han                       & 23                        & 0.529$^a$                     & 4.000$^a$                       & 0.401                 & 5.000                 & \textbf{0.69}  & 18                                                              & \textbf{0.420} & \textbf{4.593}                           & 5.21          & 2.81                                                 \\
jakobs1                   & 25                        & 0.433$^d$                     & 8.341$^a$                       & \textbf{0.366}        & \textbf{8.682}        & \textbf{0.70}  & 15                                                              & 0.288          & 9.682                                    & 2.54          & 1.65                                                 \\
jakobs2                   & 25                        & 0.418$^d$                     & 6.568$^a$                    & 0.327                 & 7.909                 & \textbf{0.77}  & 21                                                              & \textbf{0.339} & \textbf{7.568}                           & 2.84          & 1.13                                                 \\
mao                       & 20                        & 0.472$^d$                     & 3.623$^a$                       & 0.426                 & 4.000                 & \textbf{0.82}  & 20                                                              & \textbf{0.429} & \textbf{3.856}                           & 21.34         & 3.62                                                 \\
poly1a                    & 15                        & 0.456$^a$                     & 2.892$^b$                       & 0.277                 & 3.979                 & \textbf{0.92}  & 15                                                              & \textbf{0.295} & \textbf{3.629}                           & 1.67          & 0.44                                                 \\
poly2a                    & 30                        & 0.458$^a$                     & 5.782$^a$                       & 0.338                 & 6.909                 & \textbf{0.54}  & 30                                                              & \textbf{0.343} & \textbf{6.715}                           & 5.28          & 1.57                                                 \\
poly2b                    & 30                        & 0.419$^d$                     & 6.506$^b$                       & 0.314                 & 7.909                 & \textbf{0.76}  & 30                                                              & \textbf{0.321} & \textbf{7.664}                           & 5.25          & 1.93                                                 \\
poly3a                    & 45                        & 0.455$^a$                     & 8.779$^a$                       & 0.322                 & 10.909                & \textbf{0.93}  & 45                                                              & \textbf{0.324} & \textbf{10.629}                          & 16.20         & 3.11                                                 \\
poly3b                    & 45                        & 0.456$^d$                     & 8.889$^a$                       & 0.365                 & 9.951                 & \textbf{0.81}  & 43                                                              & \textbf{0.372} & \textbf{9.930}                           & 12.78         & 3.83                                                 \\
poly4a                    & 60                        & 0.453$^a$                     & 11.961$^a$                      & 0.342                 & 13.909                & \textbf{0.98}  & 60                                                              & \textbf{0.354} & \textbf{13.419}                          & 42.09         & 5.67                                                 \\
poly4b                    & 60                        & 0.431$^a$                     & 11.210$^b$                      & 0.360                 & 12.937                & \textbf{1.00}  & 58                                                              & \textbf{0.402} & \textbf{11.965}                          & 21.98         & 6.79                                                 \\
poly5a                    & 75                        & 0.455$^a$                     & 14.460$^a$                      & 0.355                 & 16.909                & \textbf{1.35}  & 75                                                              & \textbf{0.365} & \textbf{16.889}                          & 58.10         & 8.42                                                 \\
poly5b                    & 75                        & 0.479$^e$                     & 13.284$^b$                      & 0.378                 & 14.993                & \textbf{1.10}  & 73                                                              & \textbf{0.385} & \textbf{14.210}                          & 45.26         & 10.72                                                \\
shapes0                   & 43                        & 0.330$^a$                     & 11.534$^a$                      & 0.182                 & 15.714                & \textbf{0.90}  & 34                                                              & \textbf{0.233} & \textbf{13.974}                          & 21.21         & 12.91                                                \\
shapes1                   & 43                        & 0.330$^a$                     & 11.649$^a$                      & 0.182                 & 15.714                & \textbf{0.79}  & 34                                                              & \textbf{0.233} & \textbf{13.974}                          & 25.80         & 12.93                                                \\
shapes2                   & 28                        & 0.309$^a$                     & 19.727$^a$                      & 0.297                 & 19.727                & 1.08           & 28                                                              & \textbf{0.297} & \textbf{\underline{19.727}} & 5.53          & 3.66                                                 \\
shirts                    & 99                        & 0.666$^e$                     & 13.839$^b$                      & 0.409                 & 15.909                & \textbf{13.63} & 91                                                              & \textbf{0.532} & \textbf{\underline{13.769}} & 456.13        & 26.81                                                \\
swim                      & 48                        & 0.397$^a$                     & 8.771$^a$                       & 0.278                 & 10.909                & \textbf{3.62}  & 48                                                              & \textbf{0.322} & \textbf{9.982}                           & 818.79        & 167.69                                               \\
trousers                  & 64                        & 0.684$^d$                     & 4.850$^a$                       & 0.501                 & 6.000                 & \textbf{2.07}  & 64                                                              & \textbf{0.671} & \textbf{5.000}                           & 380.14        & 11.99                                                \\ \hline
\multicolumn{2}{|c|}{average}                         & \multicolumn{1}{c}{0.625} & \multicolumn{1}{c|}{8.420}  & 0.348                 & 9.887                 & \textbf{1.59}  & 37.5                                                            & \textbf{0.370} & \textbf{9.385}                           & 85.53         & 12.64                                                \\
\multicolumn{2}{|c|}{min}                             & \multicolumn{1}{c}{0.500} & \multicolumn{1}{c|}{2.390}  & 0.182                 & 3.000                 & 0.48           & 9                                                               & \textbf{0.233} & \textbf{2.870}                           & \textbf{0.38} & 0.08                                                 \\
\multicolumn{2}{|c|}{Q1}                              & \multicolumn{1}{c}{0.582} & \multicolumn{1}{c|}{4.764}  & 0.306                 & 5.500                 & \textbf{0.76}  & 19                                                              & \textbf{0.309} & \textbf{4.990}                           & 4.02          & 1.61                                                 \\
\multicolumn{2}{|c|}{Q2}                              & \multicolumn{1}{c}{0.630} & \multicolumn{1}{c|}{8.341}  & \textbf{0.355}        & \textbf{8.682}        & \textbf{0.90}  & 30                                                              & 0.354          & 9.682                                    & 16.20         & 3.62                                                 \\
\multicolumn{2}{|c|}{Q3}                              & \multicolumn{1}{c}{0.659} & \multicolumn{1}{c|}{11.592} & 0.373                 & 14.451                & \textbf{1.08}  & 53                                                              & \textbf{0.394} & \textbf{13.594}                          & 33.94         & 9.57                                                 \\
\multicolumn{2}{|c|}{max}                             & \multicolumn{1}{c}{0.831} & \multicolumn{1}{c|}{19.727} & 0.588                 & 19.727                & \textbf{13.63} & 91                                                              & \textbf{0.671} & \textbf{\underline{19.727}} & 818.79        & 167.69                                               \\ \hline
\end{tabular}
}
\end{table}%

We observe that MergeDJD consistently dominates DJD in terms of both the $F$ and $K$ metrics in all JP1 and JP2 cutting instances.
Moreover, on the two datasets, MergeDJD achieves or outperforms the best known values of $F$ in 17 out of 34 instance sets, and the BKV of $K$ in 20 out of 34 instance sets.
Given the short runtime of MergeDJD, these results are competitive with those of improvement algorithms, despite the latter requiring substantially longer runtime.
Regarding the computational time, we observe that MergeDJD can even be faster than DJD when the number of pieces after merging is significantly smaller than the original number.
This is because the reduction in the number of pieces compensates for the additional time introduced by the additional steps of MergeDJD.
In particular, when no pieces can be merged in MergeDJD 
(e.g., in the Conv I instance set), MergeDJD still outperforms DJD. 
This results demonstrates the superior effectiveness of the ICAD placement over the original CAD placement in DJD.

\begin{table}[h!]
  \centering
  \caption{Comparisons on the Nesting-MB instances. Bold values indicate the best results among the compared algorithms. Underlined values denote results that meet or outperform the best known values (BKV).}
  \label{tab:nestingresult-MB}
    \resizebox{0.8\linewidth}{!}{
    \begin{tabular}{|cc|cc|ccc|ccccc|}
\hline
\multirow{3}{*}{\begin{tabular}[c]{@{}c@{}}instance\\ set\end{tabular}} & \multirow{3}{*}{\begin{tabular}[c]{@{}c@{}}ave\\ \#pieces\end{tabular}} & \multicolumn{2}{c|}{BKV} & \multicolumn{3}{c|}{DJD} & \multicolumn{5}{c|}{MergeDJD}                                                                                                                                                        \\ \cline{3-12} 
                                                                        &                                                                         & F          & K           & F      & K       & time  & \begin{tabular}[c]{@{}c@{}}ave \#pieces \\ after merge\end{tabular} & \multicolumn{1}{c}{F} & \multicolumn{1}{c}{K} & time & \begin{tabular}[c]{@{}c@{}}merge\\ time\end{tabular} \\
\hline
albano                    & 24                        & 0.532$^d$       & 2.497$^a$      & 0.493          & 3.000          & 1.11  & 24                                                              & \textbf{0.506}                           & \textbf{2.914}                          & 10.44   & 2.62       \\
dighe1                    & 16                        & 0.374$^a$       & 1.333$^b$      & 0.288          & 2.000          & 0.80  & 15                                                              & \textbf{0.305}                           & \textbf{1.566}                          & 1.41    & 0.36       \\
dighe2                    & 10                        & 0.823$^a$       & 0.952$^b$      & 0.263          & 2.000          & 0.62  & 1                                                               & \underline{\textbf{0.823}}               & \underline{\textbf{0.952}}              & 0.53    & 0.01       \\
fu                        & 12                        & 0.478$^e$       & 3.571$^b$      & \textbf{0.442} & 3.905          & 0.74  & 12                                                              & \textbf{0.442}                           & \underline{\textbf{3.571}}              & 0.41    & 0.11       \\
han                       & 23                        & 0.420$^a$       & 2.203$^b$      & 0.326          & 3.000          & 0.70  & 15                                                              & \textbf{0.338}                           & \textbf{2.725}                          & 5.96    & 2.41       \\
jakobs1                   & 25                        & 0.579$^d$       & 3.250$^a$      & \textbf{0.522} & \textbf{4.000} & 0.75  & 12                                                              & 0.335                                   & 4.500                                   & 3.04    & 1.06       \\
jakobs2                   & 25                        & 0.513$^a$       & 3.000$^a$      & 0.384          & 4.000          & 0.51  & 19                                                              & \textbf{0.423}                           & \textbf{3.333}                          & 3.23    & 1.10       \\
mao                       & 20                        & 0.420$^a$       & 1.852$^a$      & 0.311          & 3.000          & 1.05  & 20                                                              & \underline{\textbf{0.490}}               & \textbf{1.992}                          & 42.71   & 3.76       \\
poly1a                    & 15                        & 0.349$^d$       & 1.497$^a$      & 0.296          & 2.000          & 0.47  & 15                                                              & \textbf{0.301}                           & \textbf{1.872}                          & 1.80    & 0.40       \\
poly2a                    & 30                        & 0.521$^a$       & 2.961$^a$      & 0.307          & 4.000          & 0.58  & 30                                                              & \textbf{0.340}                           & \textbf{3.652}                          & 4.08    & 1.46       \\
poly2b                    & 30                        & 0.414$^e$       & 3.420$^a$      & 0.276          & 5.000          & 0.68  & 30                                                              & \textbf{0.376}                           & \textbf{3.949}                          & 5.42    & 1.83       \\
poly3a                    & 45                        & 0.455$^a$       & 4.517$^a$      & 0.306          & 5.975          & 0.75  & 45                                                              & \textbf{0.423}                           & \textbf{5.000}                          & 10.12   & 3.33       \\
poly3b                    & 45                        & 0.462$^d$       & 4.515$^a$      & 0.322          & 6.000          & 0.54  & 42                                                              & \textbf{0.423}                           & \textbf{5.000}                          & 12.64   & 3.69       \\
poly4a                    & 60                        & 0.519$^a$       & 5.871$^a$      & 0.332          & 7.933          & 0.91  & 60                                                              & \textbf{0.395}                           & \textbf{6.821}                          & 18.98   & 5.57       \\
poly4b                    & 60                        & 0.483$^d$       & 5.751$^a$      & 0.368          & 7.000          & 0.52  & 57                                                              & \textbf{0.383}                           & \textbf{6.205}                          & 21.63   & 6.76       \\
poly5a                    & 75                        & 0.476$^a$       & 7.443$^a$      & 0.366          & 8.974          & 1.52  & 75                                                              & \textbf{0.391}                           & \textbf{8.769}                          & 59.88   & 8.58       \\
poly5b                    & 75                        & 0.488$^a$       & 6.555$^a$      & 0.377          & 7.974          & 0.82  & 72                                                              & \textbf{0.405}                           & \textbf{7.513}                          & 39.06   & 10.04      \\
shapes0                   & 43                        & 0.398$^a$       & 5.381$^a$      & 0.229          & 8.000          & 1.21  & 29                                                              & \textbf{0.297}                           & \textbf{6.524}                          & 41.44   & 10.11      \\
shapes1                   & 43                        & 0.398$^a$       & 5.381$^a$      & 0.229          & 8.000          & 1.04  & 29                                                              & \textbf{0.297}                           & \textbf{6.524}                          & 67.51   & 10.41      \\
shapes2                   & 28                        & 0.351$^e$       & 9.800$^g$      & \textbf{0.351} & {\textbf{9.800}} & 0.47  & 28                                                              & \underline{\textbf{0.351}}               & 9.933                      & 5.889   & 3.113      \\
shirts                    & 99                        & 0.666$^a$       & 6.740$^a$      & 0.603          & 7.000          & 9.77  & 91                                                              & \textbf{0.604}                           & \textbf{6.974}                          & 1010.93 & 26.92      \\
swim                      & 48                        & 0.402$^d$      & 4.411$^a$      & 0.284          & 5.981          & 3.13  & 48                                                              & \textbf{0.369}                           & \textbf{4.989}                          & 683.16  & 171.34     \\
trousers                  & 64                        & 0.590$^d$       & 2.635$^a$      & 0.372          & 4.000          & 29.05 & 64                                                              & \textbf{0.577}                           & \textbf{3.000}                          & 1259.89 & 11.73      \\ \hline
\multicolumn{2}{|c|}{average}                         & 0.483       & 4.160      & 0.350          & 5.328          & 2.51  & 36.2                                                            & \textbf{0.417}                           & \textbf{4.708}                          & 143.92  & 12.47      \\
\multicolumn{2}{|c|}{min}                             & 0.349       & 0.952      & 0.229          & 2.000          & 0.47  & 1                                                               & \textbf{0.297}                           & \underline{\textbf{0.952}}              & 0.41    & 0.01       \\
\multicolumn{2}{|c|}{Q1}                              & 0.408       & 2.566      & 0.292          & 3.452          & 0.60  & 17                                                              & \textbf{0.339}                           & \textbf{2.957}                          & 3.66    & 1.28       \\
\multicolumn{2}{|c|}{Q2}                              & 0.476       & 3.571      & 0.326          & 5.000          & 0.75  & 29                                                              & \textbf{0.391}                           & \textbf{4.500}                          & 10.44   & 3.33       \\
\multicolumn{2}{|c|}{Q3}                              & 0.520       & 5.566      & 0.375          & 7.466          & 1.08  & 52.5                                                            & \textbf{0.433}                           & \textbf{6.524}                          & 42.07   & 9.31       \\
\multicolumn{2}{|c|}{max}                             & 0.823       & 9.933      & 0.603          & \textbf{9.800} & 29.05 & 91                                                              & \underline{\textbf{0.823}}               & \underline{9.933}                       & 1259.89 & 171.34     \\ \hline
\end{tabular}
}
\end{table}

\begin{table}[h!]
  \centering
  \caption{Comparisons on the Nesting-LB instances. Bold values indicate the best results among the compared algorithms. Underlined values denote results that meet or outperform the best known values (BKV).}
  \label{tab:nestingresult-LB}
    \resizebox{0.8\linewidth}{!}{
\begin{tabular}{|cc|cc|ccc|ccccc|}
\hline
\multirow{3}{*}{\begin{tabular}[c]{@{}c@{}}instance\\ set\end{tabular}} & \multirow{3}{*}{\begin{tabular}[c]{@{}c@{}}ave\\ \#pieces\end{tabular}} & \multicolumn{2}{c|}{BKV} & \multicolumn{3}{c|}{DJD} & \multicolumn{5}{c|}{MergeDJD}                                                                                                                                                        \\ \cline{3-12} 
                                                                        &                                                                         & F          & K           & F      & K       & time  & \begin{tabular}[c]{@{}c@{}}ave \#pieces \\ after merge\end{tabular} & \multicolumn{1}{c}{F} & \multicolumn{1}{c}{K} & time & \begin{tabular}[c]{@{}c@{}}merge\\ time\end{tabular} \\
\hline
albano                    & 24                        & 0.416$^a$       & 1.374$^a$      & 0.393 & 2.000 & 1.40  & 24                                                              & \textbf{0.396}          & \textbf{1.627} & 34.66   & 2.40       \\
dighe1                    & 16                        & 0.329$^a$       & 0.707$^a$      & 0.329 & 1.000          & 0.50  & 15                                                              & \underline{\textbf{0.329}} & \textbf{0.962} & 0.80    & 0.33       \\
dighe2                    & 10                        & 0.260$^a$       & 0.587$^b$      & 0.260 & 1.000          & 0.38  & 1                                                               & \underline{\textbf{0.260}} & \textbf{0.714} & 0.36    & 0.04       \\
fu                        & 12                        & 0.526$^a$       & 1.571$^a$      & 0.485          & 2.000          & 0.49  & 11                                                              & \textbf{0.494} & \textbf{1.875} & 0.31    & 0.04       \\
han                       & 23                        & 0.412$^d$       & 1.217$^b$      & 0.255          & 2.000          & 0.50  & 15                                                              & \textbf{0.297} & \textbf{1.543} & 4.67    & 2.97       \\
jakobs1                   & 25                        & 0.617$^d$       & 1.763$^b$      & \textbf{0.595} & \textbf{2.000} & 0.50  & 11                                                              & 0.342          & 2.375          & 3.49    & 1.18       \\
jakobs2                   & 25                        & 0.464$^a$       & 1.686$^a$      & 0.444          & 2.000 & 0.44  & 19                                                              & \textbf{0.445} & \textbf{2.000} & 2.28    & 1.09       \\
mao                       & 20                        & 0.610$^a$       & 0.961$^a$      & 0.209          & 2.000          & 0.66  & 20                                                              & \textbf{0.278} & \textbf{1.129} & 15.77   & 3.68       \\
poly1a                    & 15                        & 0.368$^a$       & 0.861$^a$      & 0.367 & 2.000          & 0.54  & 15                                                              & \textbf{0.368} & \textbf{1.000} & 1.31    & 0.43       \\
poly2a                    & 30                        & 0.394$^e$       & 1.632$^a$      & 0.224          & 3.000          & 0.56  & 30                                                              & \textbf{0.371} & \textbf{2.000} & 3.15    & 1.49       \\
poly2b                    & 30                        & 0.459$^d$       & 1.870$^a$      & 0.240          & 3.000          & 0.49  & 30                                                              & \textbf{0.281} & \textbf{2.385} & 5.27    & 1.81       \\
poly3a                    & 45                        & 0.416$^d$       & 2.452$^a$      & 0.268          & 4.000          & 0.77  & 45                                                              & \textbf{0.375} & \textbf{2.960} & 9.49    & 3.19       \\
poly3b                    & 45                        & 0.418$^d$       & 2.448$^a$      & 0.370          & 3.000          & 0.44  & 42                                                              & \textbf{0.381} & \textbf{2.819} & 11.55   & 3.80       \\
poly4a                    & 60                        & 0.431$^a$       & 3.274$^a$      & 0.286          & 5.000          & 0.94  & 60                                                              & \textbf{0.381} & \textbf{3.961} & 12.09   & 5.61       \\
poly4b                    & 60                        & 0.415$^a$       & 3.231$^a$      & 0.352          & 4.000          & 0.62  & 57                                                              & \textbf{0.364} & \textbf{3.646} & 24.04   & 6.91       \\
poly5a                    & 75                        & 0.437$^a$       & 4.231$^a$      & 0.303          & 6.000          & 0.64  & 75                                                              & \textbf{0.383} & \textbf{4.963} & 67.04   & 8.57       \\
poly5b                    & 75                        & 0.476$^a$       & 3.742$^a$      & 0.342          & 5.000          & 0.69  & 71                                                              & \textbf{0.464} & \textbf{4.000} & 34.32   & 9.80       \\
shapes0                   & 43                        & 0.462$^a$       & 3.000$^a$      & 0.260          & 5.000          & 0.83  & 21                                                              & \textbf{0.281} & \textbf{3.679} & 86.00   & 8.53       \\
shapes1                   & 43                        & 0.462$^a$       & 3.000$^a$      & 0.260          & 5.000          & 0.89  & 21                                                              & \textbf{0.281} & \textbf{3.679} & 110.97  & 8.37       \\
shapes2                   & 28                        & 0.485$^c$       & 4.700$^b$      & \textbf{0.437} & \textbf{4.800} & 0.65  & 24                                                              & 0.435          & 4.910          & 10.19   & 3.31       \\
shirts                    & 99                        & 0.652$^d$       & 3.916$^a$      & 0.588          & 4.000 & 10.52 & 91                                                              & \textbf{0.591} & \textbf{4.000} & 1849.41 & 26.95      \\
swim                      & 48                        & 0.416$^d$       & 2.414$^a$      & 0.238          & 2.998          & 1.85  & 48                                                              & \textbf{0.335} & \textbf{2.829} & 1048.29 & 166.73     \\
trousers                  & 64                        & 0.462$^d$       & 1.405$^a$      & 0.408          & 2.000          & 21.09 & 64                                                              & \textbf{0.451} & \textbf{1.926} & 2375.45 & 12.08      \\ \hline
\multicolumn{2}{|c|}{average}                         & 0.452       & 2.263      & 0.344          & 3.165          & 2.02  & 35.2                                                            & \textbf{0.373} & \textbf{2.651} & 248.30  & 12.14      \\
\multicolumn{2}{|c|}{min}                             & 0.260       & 0.587      & 0.209          & 1.000          & 0.38  & 1                                                               & \underline{\textbf{0.260}} & \textbf{0.714} & 0.31    & 0.04       \\
\multicolumn{2}{|c|}{Q1}                              & 0.416       & 1.390      & 0.260          & 2.000          & 0.50  & 17                                                              & \textbf{0.313} & \textbf{1.751} & 3.32    & 1.33       \\
\multicolumn{2}{|c|}{Q2}                              & 0.437       & 1.870      & 0.329          & 3.000          & 0.64  & 24                                                              & \textbf{0.371} & \textbf{2.385} & 11.55   & 3.31       \\
\multicolumn{2}{|c|}{Q3}                              & 0.470       & 3.116      & 0.401          & 4.400          & 0.86  & 52.5                                                            & \textbf{0.416} & \textbf{3.679} & 50.85   & 8.45       \\
\multicolumn{2}{|c|}{max}                             & 0.652       & 4.700      & \textbf{0.595} & 6.000          & 21.09 & 91                                                              & 0.591          & \textbf{4.963} & 2375.45 & 166.73     \\ \hline
\end{tabular}
}
\end{table}

On the SB, MB, and LB nesting instances, we observe that MergeDJD outperforms DJD in terms of the F metric in 61 out of 69 instances. Among these, MergeDJD achieves or outperforms the best known value (BKV) of F in 5 instances.
Regarding the K metric, MergeDJD outperforms DJD in 63 out of 69 instances, and achieves or outperforms the BKV of K in 5 instances.
In general, only a small number of pieces are merged in these instances; nevertheless, MergeDJD consistently provides improvements over DJD, indicating that the proposed enhancements are effective even when limited merging is possible.
We also observe that the runtime of MergeDJD increases significantly for instances with a large number of pieces. This is because ICAD introduces the points of placed pieces as candidate initial positions, whose number can grow exponentially with respect to the number of placed pieces. As a result, the time required to verify the feasibility of each initial position increases substantially.
A possible way to alleviate this issue is to sample only a proportion of the candidate initial positions.

\begin{figure}[h!]
    \centering
    \includegraphics[width=0.5\linewidth]{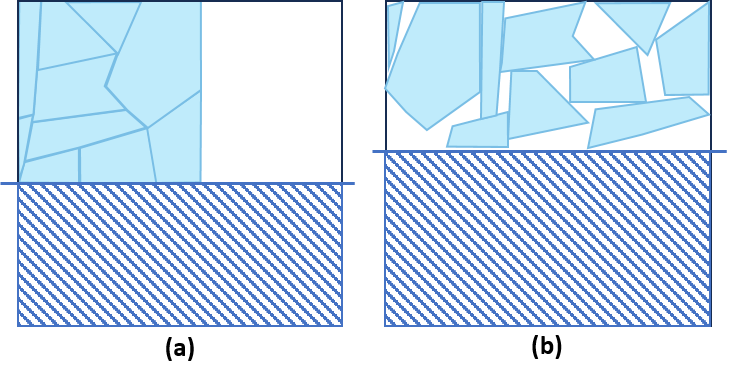}
    \caption{A counterintuitive scenario. Given a large bin, $F$ metric fails to distinguish the two placements (a) and (b), while $K$ metric prioritizes less compact placement (b).}
    \label{fig:counterintuitive}
\end{figure}
As the bin size grows, we found that more pieces are merged. 
Specifically, the pieces of the dighe2 instance in the MB and LB sets are merged into a single rectangle.
As shown in Figure~\ref{fig:counterintuitive} (a), the combined piece represents the most compact placement of these pieces.
However, compared with a less compact placement (as shown in   Figure~\ref{fig:counterintuitive} (b)), the F metric remains unchanged, while the less compact placement yields a better K metric.
This counterintuitive observation suggests that the current evaluation metrics may not adequately capture placement compactness, indicating the need for more appropriate evaluation criteria.

\begin{figure}[h!]
    \centering
    \includegraphics[width=1\linewidth]{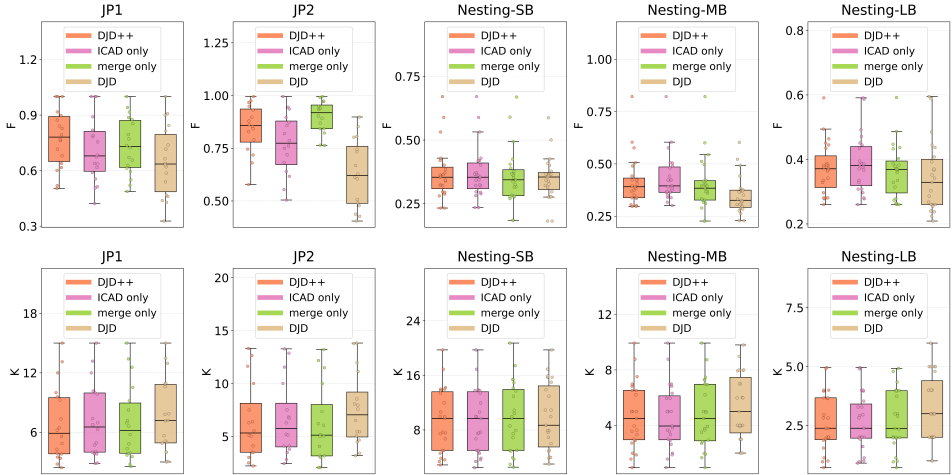}
    \caption{The impact of the improvements of MergeDJD.}
    \label{fig:ablation}
\end{figure}

\subsection{Effectiveness of the New Components}
In this section, we analyze the effectiveness of the new components introduced in  MergeDJD.
As shown in Figure~\ref{fig:ablation}, we compare MergeDJD with three ablated variants, namely {\em merge-only}, {\em ICAD-only} and original DJD, in terms of the F and K metrics across all instance groups. 

The results suggest that, compared to DJD, the two improvements introduced in MergeDJD have a positive impact on both F and K in all instances. 
Furthermore, we observe that merging is the primary source of performance improvement in MergeDJD, while ICAD mainly enhances robustness and overall effectiveness. 
Specifically, on JP1, MergeDJD clearly outperforms all variants in terms of the F metric, achieving higher median values and reduced variance. On JP2, the merge-only variant attains F values comparable to, and in some cases slightly higher than, those of MergeDJD, indicating that the benefit of ICAD is less pronounced in this data set.
For nesting instances (SB, MB and LB), MergeDJD consistently achieves equal or better F values than the three variants, although the performance gaps are relatively small due to the limited number of pieces that can be merged.
Regarding the K metric, MergeDJD generally achieves lower median values than the three variants in most of the instance groups. An exception is observed in Nesting-MB instances, where the ICAD-only variant attains slightly better K values than MergeDJD.

\section{Conclusion}
\label{sec:conclusion}
This paper presented MergeDJD, a new constructive heuristic for the two-dimensional irregular bin packing problem.
MergeDJD extends the DJD algorithm by introducing a merge strategy and an improved placement heuristic, ICAD.
The merging strategy combines compatible pieces into combined pieces.
ICAD improves the placement of non-convex and merged pieces.
Experimental results on standard benchmark instances show that MergeDJD consistently outperforms DJD in short runtimes.
Ablation studies indicate that the merge strategy is the main contributor to the observed improvements, while ICAD further enhances robustness and solution quality.

Future work may explore several directions.
Learning-based methods can be used to guide merging or placement decisions. For example, they can predict promising merge candidates or ranking placement positions based on geometric features. This can reduce unnecessary trials while preserving the constructive nature of the algorithm.
Parallel implementations of MergeDJD may further reduce runtime for large-scale applications. For instance,  multiple merge candidates or placement positions can be evaluated concurrently for large instances.

\section*{Acknowledgments}
This work is supported by the National Natural Science Foundation of China under grant 62372093 and the Shenzhen Science and Technology Program under grant KJZD20240903095712016.
The authors thank Zibo Zhou for his help with an early prototype implementation.


\bibliography{cas-refs}
\end{document}